\documentclass{aa}  
\usepackage{natbib}
\usepackage{graphicx}
\usepackage{txfonts}
\usepackage{lscape}
\begin{document}
   \title{Quantitative spectroscopy of Galactic BA-type supergiants\thanks{Based 
   on observations collected at the Centro Astron\'omico\,Hispa\-no
   Alem\'an at Calar Alto (CAHA), operated jointly
   by the Max-Planck Institut f\"ur Astronomie and the Instituto de
   Astrof\'isica de Andaluc\'ia (CSIC), proposals H2001-2.2-011
   and H2005-2.2-016.}\fnmsep\thanks{Based on
   observations obtained at the European Southern Obser\-vatory,
   proposals 62.H-0176 and 079.B-0856(A). 
   Additional data were adopted from the UVES Paranal Observatory Project 
   (ESO DDT Program ID 266.D-5655).}}
   \subtitle{I. Atmospheric parameters}
   \author{M. Firnstein
          \and
          N. Przybilla
          }

   \institute{Dr. Karl Remeis-Sternwarte \& ECAP, 
	      Universit\"at Erlangen-N\"urnberg,
	      Sternwart\-str.~7, D-96049 Bamberg, Germany\\
              \email{Markus.Firnstein@sternwarte.uni-erlangen.de;Norbert.Przybilla@sternwarte.uni-erlangen.de}
                           }

   \date{Received ; accepted }

 
  \abstract
   {BA-type supergiants show a high potential as versatile indicators for 
   modern astronomy. 
   This paper constitutes the first in a series that aims at a
   systematic spectroscopic study of Galactic BA-type supergiants. 
   Various problems will be addressed, including in particular
   observational constraints on the evolution of 
   massive stars and a determination of abundance gradients in the Milky Way.}
   {The focus here is on the determination of accurate and precise 
   atmospheric parameters for a sample of Galactic BA-type supergiants
   as prerequisite for all further analysis. Some first 
   applications include a recalibration of functional relationships
   between spectral-type, intrinsic colours, bolometric corrections and 
   effective
   temperature, and an exploration of the reddening-free Johnson $Q$ and
   Str\"omgren $[c_1]$ and $\beta$-indices as photometric indicators for 
   effective temperatures and gravities of BA-type supergiants.}
   {An extensive grid of theoretical spectra is computed based on a
   hybrid non-LTE approach, covering the relevant parameter space in effective 
   temperature, surface gravity, helium abundance, microturbulence and elemental 
   abundances.
   The atmospheric parameters are derived spectroscopically by line-profile fits 
   of our theoretical models to high-resolution and high-S/N spectra
   obtained at various observatories. Ionization equilibria of
   multiple metals and 
   the Stark-broadened hydrogen and the neutral helium lines constitute
   our primary indicators for the parameter determination,
   supplemented by (spectro-)photometry from the UV to the near-IR.}
   {We obtain accurate atmospheric parameters for 35 sample supergiants 
   from a homogeneous analysis. Data on effective temperatures, surface
   gravities, helium abundances, microturbulence, macroturbulence and
   rotational velocities are presented. The interstellar reddening and the
   ratio of total-to-selective extinction towards the stars are determined. 
   Our empirical spectral-type--$T_{\rm eff}$ scale is steeper than
   reference relations from the literature, the stars are
   significantly bluer than usually assumed, and bolometric 
   corrections differ significantly from established literature values. 
   Photometric $T_{\rm eff}$-determinations 
   based on the reddening-free $Q$-index are found to be of limited use for 
   studies of BA-type supergiants because of large errors of typically
   $\pm$5\%\,(1$\sigma$ statistical)$\pm$3\%\,(1$\sigma$ systematic),
   compared to a spectroscopically achieved precision of 1-2\%
   (combined statistical and systematic uncertainty with our
   methodology).
   The reddening-free $[c_1]$-index and $\beta$ on the other hand
   are found to provide useful starting values for 
   high-precision/accuracy analyses, with uncertainties of 
   $\pm$1\%\,$\pm$2.5\% in $T_{\rm eff}$,
   and $\pm$0.04$\pm$0.13\,dex in $\log g$ \,(1$\sigma$-statistical, 
   1$\sigma$-systematic,~respectively).} 
   {}


   \titlerunning{Quantitative Spectroscopy of Galactic BA-type
   Supergiants.~I}

   \maketitle
%

\section{Introduction}
Supergiants of late B and early A-type (BA-type supergiants) are among 
the visually brightest stars, reaching absolute magnitudes of up to 
$M_{\mathrm{V}}$\,$\approx$\,$-$9.5. Therefore, they show a high potential 
as versatile indicators for stellar and galactic studies over large distances 
using ground-based telescopes of the 8-10\,m class. The focus of quantitative studies of 
individual BA-type supergiants lies in extragalactic research at
present. Objects in many of the star-forming galaxies of the Local Group 
and even beyond have been investigated at high and intermediate 
spectral resolution \citep[see e.g.~the reviews by][and references
therein]{Vennetal03,kudritzki08b,Kudritzki10}.
Such observations in other galaxies allow the spatial distribution of 
elemental abundances to be mapped, using individual supergiants as
tracers. This can provide observational
constraints on galactochemical evolution models of different galaxy
types as well as on
stellar evolution models for a wide range of metallicities, see
e.g. \citet{przybilla08} for a review, and in a broader context on the 
galaxy mass--metallicity relationship \citep{Kudritzki12}.

Moreover, BA-type supergiants can act as standard candles for distance 
determinations via use of the wind momentum-luminosity relationship 
\citep[][]{Puls,Kudritzki99} or the flux-weighted gravity-luminosity 
relationship \citep[FGLR,][]{Kudritzki03,Kudritzki08}. The latter is of particular 
interest as the stellar metallicity and the interstellar reddening are
also derived in the quantitative analysis. In consequence, sources of 
systematic errors that trouble photometric indicators like the
Cepheid period-luminosity relationship \citep{freedman01,Kudritzki08}
can be constrained by this spectroscopic method.
Systematic studies therefore promise refinements to the extragalactic distance
scale and to the determination of the Hubble constant to be achieved
\citep{KudUrb12}.

Accordingly, detailed quantitative studies of {\em nearby} stars of
this kind, in the Milky Way,
should be of great interest. Modern spectrographs open up the
possibility to obtain spectra of exceptional quality of these bright
stars, even with small telescopes of the 2\,m-class. Galactic targets 
may therefore act as testbeds for scrutinising stellar atmosphere analysis
techniques for BA-type supergiants before using them in 
extragalactic applications \citep{NorbertPhD,Norbert}. However, there is much
more to gain from a systematic study of BA-type supergiants in the
Milky Way. New light may also be shed on some facets of the actions of
the cosmic cycle of matter in the high-metallicity environment of a typical giant 
spiral galaxy.

Studies of stellar structure and evolution are the basis for
our understanding of the synthesis of the heavy elements (metal yields)
and their injection into the cosmic matter cycle by stellar winds and
supernovae. The most recent generations of evolution models 
for massive stars that account for effects of rotation and mass loss 
as well as magnetic fields 
\citep[see e.g. the recent review by][and references therein]{MaMe12}
are highly successful in describing many observational aspects of the
massive star populations in general. However, many details -- in
particular related to the subtle signatures of mixing of CNO-cycled
products in the stars -- are subject to intense debate at
present, see e.g. \cite{Hunter}, \cite{Maeder09} and \cite{CNOmix}. Here, highly accurate
observational data on fundamental stellar parameters and light element
abundances are required for thoroughly testing the models. 
BA-type supergiants are primary candidates for the study of the
post-main-sequence evolution, linking core H-burning main-sequence stars with 
core He-burning red supergiants (some He-burning objects may also be
on a blue loop). 

Furthermore, studies of stars are the key to understand the chemical
evolution of the Galaxy. Luminous BA-type supergiants are highly 
useful in mapping elemental abundance patterns throughout the Milky Way. 
Systematic investigations of these short-lived stars can constrain the
present-day end point of $\sim$13\,Gyr of Galactic evolution in form of 
abundance gradients throughout the Galactic disk. These represent one
of the major observational constraints to Galactochemical evolution models. 
Previous work using \ion{H}{ii}-regions \citep[e.g.][and references
therein]{Shaver83,Esteban05,Rudolph06}, B-type main sequence stars
\citep[e.g.][]{Gummersbach98,Rolleston00,Daflon04}
or Cepheids \citep[e.g.][and references therein]{Andrievsky04,Pedicelli09} 
may be verified, complemented and extended by studies of BA-type supergiants.
Precise and unbiased elemental abundances for a larger sample of
objects with well-constrained 
distances, distributed throughout the disk, are required for this.

While the literature on properties of hot, massive stars of O- and
early B-type
is rich, few studies have concentrated on tepid BA-type supergiants in the past
$\sim$20 years, where model atmosphere techniques have reached some
kind of maturity. Pioneering work was done by
\citet{Venn95a,Venn95b}, later updated by \citet{Venn03}, who determined 
atmospheric parameters and chemical 
abundances for a sample of 22 Galactic A-type supergiants, mostly less-luminous 
ones of luminosity class II and Ib. \citet{Verdugo}
derived basic stellar parameters for 31 early A-type supergiants,
while the compilation of \citet{Lyubimkov} provides such data for 8 mostly
less-luminous late A-type supergiants. Moreover, atmospheric parameters for a smaller
number of late B-type supergiants were provided by \citet{McErlean}
and \citet{Fraser}. Finally, the work of \citet[][and references
therein]{Takeda} on a representative sample of BA-type supergiants
is the most comprehensive one concerning abundance
determinations that account for deviations from the standard
assumption of local thermodynamic equilibrium (non-LTE) so far.

However, a comprehensive and homogeneous study of Galactic BA-type supergiants in
which atmospheric and fundamental stellar parameters, and abundances 
for light, $\alpha$-process and iron group elements alike are derived
is still unavailable. This was our motivation to embark on the present
work, combining state-of-the-art observational data based on
echelle spectra with a sophisticated hybrid non-LTE analysis methodology
\citep[with extensions to facilitate a semi-automatic analysis as
described below]{Norbert}. 
The aim is to provide as accurate (i.e. minimised systematic
uncertainties) and precise (i.e. minimised statistical errors) 
data as presently possible for a larger number of 
objects. Here, in Paper\,I, we introduce the observational database
and derive the atmospheric parameters of the sample stars. This
constitutes the basis for all further discussion, that will
concentrate on observational constraints to evolution models for
massive stars (Paper\,II) and to models of Galactochemical evolution (Paper\,III).

The paper is organised as follows. In Sect.~2 we describe the available 
observational data, while our methods for the determination of 
stellar parameters and elemental abundances are discussed in Sect.~3. 
We present the results of our studies in Sect.~4 and some first applications
in Sect.~5. Finally, a short summary is given in Sect.~6.


\begin{table*}[ht]
\centering
 \setlength{\tabcolsep}{.1cm}
\caption{The star sample: id, spectral type, OB association membership\tablefootmark{a}, photometry\tablefootmark{b}, and observational details\tablefootmark{c}. }
\label{observations1}
 \begin{tabular}{r@{\hspace{2mm}}lllc@{\hspace{5mm}}r@{$\pm$}lr@{$\pm$}lr@{$\pm$}llrr}
 \noalign{}
\hline\hline
\#  & Object  & Sp.\,T\tablefootmark{d} & Sp.\,T\tablefootmark{e} & OB\,Assoc.
&\multicolumn{2}{c}{$V$}&\multicolumn{2}{c}{$B-V$}&\multicolumn{2}{c}{$U-B$}& Date & $T_\mathrm{exp}$ & $S/N_V$\\
    &         &                         &                         &
    &\multicolumn{2}{c}{mag}&\multicolumn{2}{c}{mag}&\multicolumn{2}{c}{mag}&
    & s\\

\hline\\[-3mm]
{\sc Foces}~&$R$\,=\,40\,000\\
\hline\\[-2mm]
1  & \object{HD12301}  & B8\,Ib\tablefootmark{f}& B8\,Ib    & Field    & 5.589 & 0.011 &  0.370 & 0.014 & $-$0.276 & 0.009 & 30/09/2001  & 480     &    203	\\
2  & \object{HD12953}  & A1\,Iae\tablefootmark{f}& A1\,Iae  & Per\,OB1 & 5.691 & 0.021 &  0.614 & 0.007 & $-$0.014 & 0.009 & 26/09/2001  & 300     &    230	\\
3  & \object{HD13476}  & A3\,Iab   & A3\,Iab   & Per\,OB1 & 6.431 & 0.020 &  0.600 & 0.013 &  0.220 & 0.028 & 30/09/2001  & 900     &    202	\\
4  & \object{HD13744}  & A0\,Iab   & A0\,Iab   & Per\,OB1 & 7.592 & 0.014 &  0.741 & 0.012 &  0.180 & 0.000 & 27/09/2005  & 2700    &    182	\\
5  & \object{HD14433}  & A1\,Ia\tablefootmark{f}& A1\,Ia    & Per\,OB1 & 6.401 & 0.019 &  0.567 & 0.008 &  0.030 & 0.010 & 30/09/2001  & 600     &    217	\\
6  & \object{HD14489}  & {\em A2\,Ia}\tablefootmark{f}& A1\,Iab& Per\,OB1 & 5.178 & 0.009 &  0.369 & 0.004 & $-$0.110 & 0.000 & 26/09/2001  & 360     &    246	\\
7  & \object{HD20041}  & A0\,Ia    & A0\,Ia    & Cam\,OB1 & 5.795 & 0.019 &  0.712 & 0.020 &\multicolumn{2}{c}{0.090}& 30/09/2001  & 600     &  234	\\
8  & \object{HD21291}  & B9\,Ia\tablefootmark{f}& B9\,Ia    & Cam\,OB1 & 4.213 & 0.019 &  0.412 & 0.008 & $-$0.234 & 0.009 & 26/09/2001  & 2$\times$240   &  259	\\
9  & \object{HD39970}  & A0\,Ia    & A0\,Ia    & Field    & 6.018 & 0.004 &  0.386 & 0.005 & $-$0.192 & 0.010 & 30/09/2001  & 600     &    270	\\
10 & \object{HD46300}  & A0\,Ib\tablefootmark{f}& A0\,Ib    & Mon\,OB1 & 4.498 & 0.008 &  0.007 & 0.009 & $-$0.217 & 0.041 & 29/09/2005  & 180     &    206	\\
11 & \object{HD186745} & B8\,Ia    & B8\,Ia    & Vul\,OB1 & 7.030 & 0.008 &  0.930 & 0.002 &  0.028 & 0.007 & 25/09/2001  & 900     &    163	\\
12 & \object{HD187983} & A1\,Ia    & A1\,Ia    & Field    & 5.590 & 0.026 &  0.684 & 0.017 &  0.173 & 0.149 & 25/09/2001  & 300     &    187	\\
13 & \object{HD197345} & A2\,Ia\tablefootmark{g}& A2\,Ia    & Cyg\,OB7 & 1.246 & 0.015 &  0.092 & 0.007 & $-$0.233 & 0.008 & 21/09/2005  & 8$\times$20    &    798	\\
14 & \object{HD202850} & B9\,Iab\tablefootmark{f}& B9\,Iab  & Cyg\,OB4 & 4.233 & 0.009 &  0.123 & 0.011 & $-$0.386 & 0.026 & 29/09/2001  & 120     &    231	\\
15 & \object{HD207260} & A2\,Iab\tablefootmark{f}& A2\,Iab  & Cep\,OB2 & 4.289 & 0.007 &  0.518 & 0.011 &  0.119 & 0.018 & 26/09/2001  & 120     &    370	\\
16 & \object{HD207673} & A2\,Ib\tablefootmark{f}& A2\,Ib    & Field    & 6.467 & 0.005 &  0.410 & 0.000 &\multicolumn{2}{c}{0.060}& 29/09/2001  & 720     &  195	\\
17 & \object{HD208501} & B8\,Ib\tablefootmark{f}& B8\,Ib    & Cep\,OB2 & 5.796 & 0.004 &  0.724 & 0.008 & $-$0.022 & 0.007 & 26/09/2001  & 480     &    231	\\
18 & \object{HD210221} & A3\,Ib\tablefootmark{f}& A3\,Ib    & Field    & 6.140 & 0.000 &  0.414 & 0.017 &  0.240 & 0.000 & 26/09/2001  & 720     &    271	\\
19 & \object{HD212593} & {\em B9\,Iab}\tablefootmark{f}& B9\,Ib  & Field    & 4.569 & 0.018 &  0.086 & 0.004 & $-$0.342 & 0.006 & 29/09/2001  & 2$\times$180   &  403	\\
20 & \object{HD213470} & {\em A3\,Ia}    & A3\,Iab   & Cep\,OB1 &\multicolumn{2}{c}{6.650}&\multicolumn{2}{c}{0.560}&\multicolumn{2}{c}{0.240} & 29/09/2001  & 900     &  249	\\
21 & \object{BD+602582}& B8\,Iab   & B8\,Iab   & Cas\,OB2 & 8.694 & 0.261 &  0.770 & 0.016 &  0.017 & 0.011 & 27/09/2001  & 2400    &    140	\\
22 & \object{HD223960} & {\em A0\,Ia}\tablefootmark{f}& B9\,Ia& Cas\,OB5 & 6.895 & 0.009 &  0.715 & 0.009 & $-$0.050 & 0.047 & 25/09/2001  & 1200    &    226	\\[1mm]
\hline\\[-3mm]
{\sc Foces}~&$R$\,=\,65\,000\\
\hline\\[-2mm]
23 & \object{HD195324} & A1\,Ib    & A1\,Ib    & Field    & 5.880 & 0.000 &  0.524 & 0.014 &\multicolumn{2}{c}{0.100}& 07/10/2001  & 2$\times$1000  &  618     \\[1mm]
\hline\\[-3mm]
{\sc Feros}~&$R$\,=\,48\,000 \\
\hline\\[-2mm]
24 & \object{HD34085}  & B8\,Ia\tablefootmark{g}& B8\,Ia    & Ori\,OB1 & 0.138 & 0.032 & $-$0.029 & 0.004 & $-$0.666 & 0.018 & 14/11/1998  & 20      &  634	\\
25 & \object{HD87737}  & A0\,Ib\tablefootmark{g}& A0\,Ib    & Field    & 3.486 & 0.053 & $-$0.026 & 0.015 & $-$0.206 & 0.028 & 21/01/1999  & 120     &    440	\\
26 & \object{HD91533}  & A2\,Iab   & A2\,Iab   & Car\,OB1 & 6.005 & 0.019 &  0.318 & 0.011 & $-$0.075 & 0.034 & 23/05/2007  & 100     &    229	\\
27 & \object{HD111613} & {\em A2\,Iabe}  & A1\,Ia    & Cen\,OB1 & 5.741 & 0.019 &  0.384 & 0.022 & $-$0.088 & 0.026 & 23/01/1999  & 600     &    376	\\
28 & \object{HD149076} & {\em B8\,Iab}   & B9\,Ib    & Ara\,OB1b& 7.373 & 0.018 &  0.485 & 0.009 & $-$0.118 & 0.021 & 24/05/2007  & 280     &    230	\\
29 & \object{HD149077} & {\em B9\,Ib}    & A0\,Ib    & Ara\,OB1a& 7.433 & 0.082 &  0.470 & 0.021 &  0.097 & 0.049 & 24/05/2007  & 310     &    261	\\
30 & \object{HD165784} & {\em A2/A3\,Iab}& A2\,Iab   & Sgr\,OB1 & 6.538 & 0.016 &  0.856 & 0.005 &  0.279 & 0.056 & 09/07/2007  & 140     &    145	\\
31 & \object{HD166167} & {\em B9.5\,Ib}  & A0\,Ib    & Sgr\,OB1 & 8.605 & 0.009 &  0.560 & 0.000 &  0.036 & 0.047 & 09/07/2007  & 610     &    117	\\[1mm]
\hline\\[-3mm]
{\sc Uves}~&$R$\,=\,80\,000 \\
\hline\\[-2mm]
32 & \object{HD80057}  & {\em A1\,Ib}    & A1\,Iab   & Vela\,OB1\tablefootmark{h}& 6.044 & 0.016 &  0.285 & 0.006 & $-$0.117 & 0.021 & 24/02/2003  & 2$\times$50    &    293	\\
33 & \object{HD102878} & A2\,Iab   & A2\,Iab   & Cru\,OB1 & 5.695 & 0.017 &  0.265 & 0.009 & $-$0.119 & 0.057 & 06/01/2002  & 54$+$139  &  442	\\
34 & \object{HD105071} & {\em B8\,Ia/Iab}& B8\,Iab   & Field    & 6.316 & 0.024 &  0.200 & 0.010 & $-$0.436 & 0.034 & 26/02/2002  & 2$\times$74    &    382	\\
35 & \object{HD106068} & {\em B8\,Ia/Iab}& B8\,Iab   & Field    & 5.920 & 0.010 &  0.297 & 0.007 & $-$0.284 & 0.151 & 20/07/2001  & 2$\times$53    &    415	\\[1mm]
\hline\\[-6mm]
 \end{tabular}
\tablefoot{                                                                        
\tablefoottext{a}{\citet{blaha};}
\tablefoottext{b}{\citet{mermilliod};}
\tablefoottext{c}{note that the exposure times for UVES objects can vary in different wavelength bands;}
\tablefoottext{d}{adopted from the SIMBAD database at CDS, set in italics if a revision appears to be required (see next column);}
\tablefoottext{e}{this work;}
\tablefoottext{f}{MK standards from \citet{JoMo53};}
\tablefoottext{g}{anchor points of the MK system \citep{Garrison94};}
\tablefoottext{h}{\citet{vela}.}
}
\end{table*}

\section{Observations and data reduction}
While high-resolution and high signal-to-noise (S/N) spectra are required in order to achieve 
the high accuracy and precision we aim for, a near-complete wavelength coverage 
from $\sim$3900 to 9100\,{\AA} is also essential for our analysis. 
Only then all the strategic sets of lines both for the parameter and the
abundance analysis are available in all the stars.

We obtained data on 35 bright stars ($V$\,$<$\,8.7\,mag) using three
echelle spectrographs both on the northern and southern 
hemisphere, thus sampling a wide range in Galactic longitude.
The targets were chosen in a way that covers the examined parameter domain 
(B8 to A3 in spectral type, Ib to Ia in luminosity class, according to
an initial classification from the SIMBAD database at CDS\footnote{\tt
http://simbad.u-strasbg.fr/simbad/}) rather homogeneously. 
However, overall more luminous objects were favoured in order to
facilitate larger distances to be reached within our study. Stars in both 
OB associations (25~objects) and in the field (10~objects) were observed. 
The resulting list of program stars, their spectral classification,
association membership, photometric properties and observational
details (observation dates, exposure times and S/N-ratio of the
spectra) are summarised 
in Table~\ref{observations1}. Note that objects 
in common with \citet{Norbert} and \citet{Schiller} were reanalysed for the 
sake of homogeneity (see that papers on details of the data reduction).

Most of the spectra were obtained in observing runs with the Fibre 
Optics Echelle Cassegrain Spectrograph \citep[{\sc Foces},][]{Pfeiffer} on the 
Calar Alto~2.2\,m telescope in 2001 and 2005. These cover a wavelength 
range from 3860 to 9580\,{\AA} at a resolving power 
$R$\,$=$\,$\lambda / \Delta \lambda$\,$\approx$\,40\,000. 
Only the data for HD\,195324 were acquired with a different setup, 
trading higher resolution for a lower wavelength coverage. Relatively bright 
objects were observed in order to reach the desired spectrum quality of more 
than 150 in S/N ratio.

\begin{figure}
\centering
\includegraphics[width=.82\linewidth]{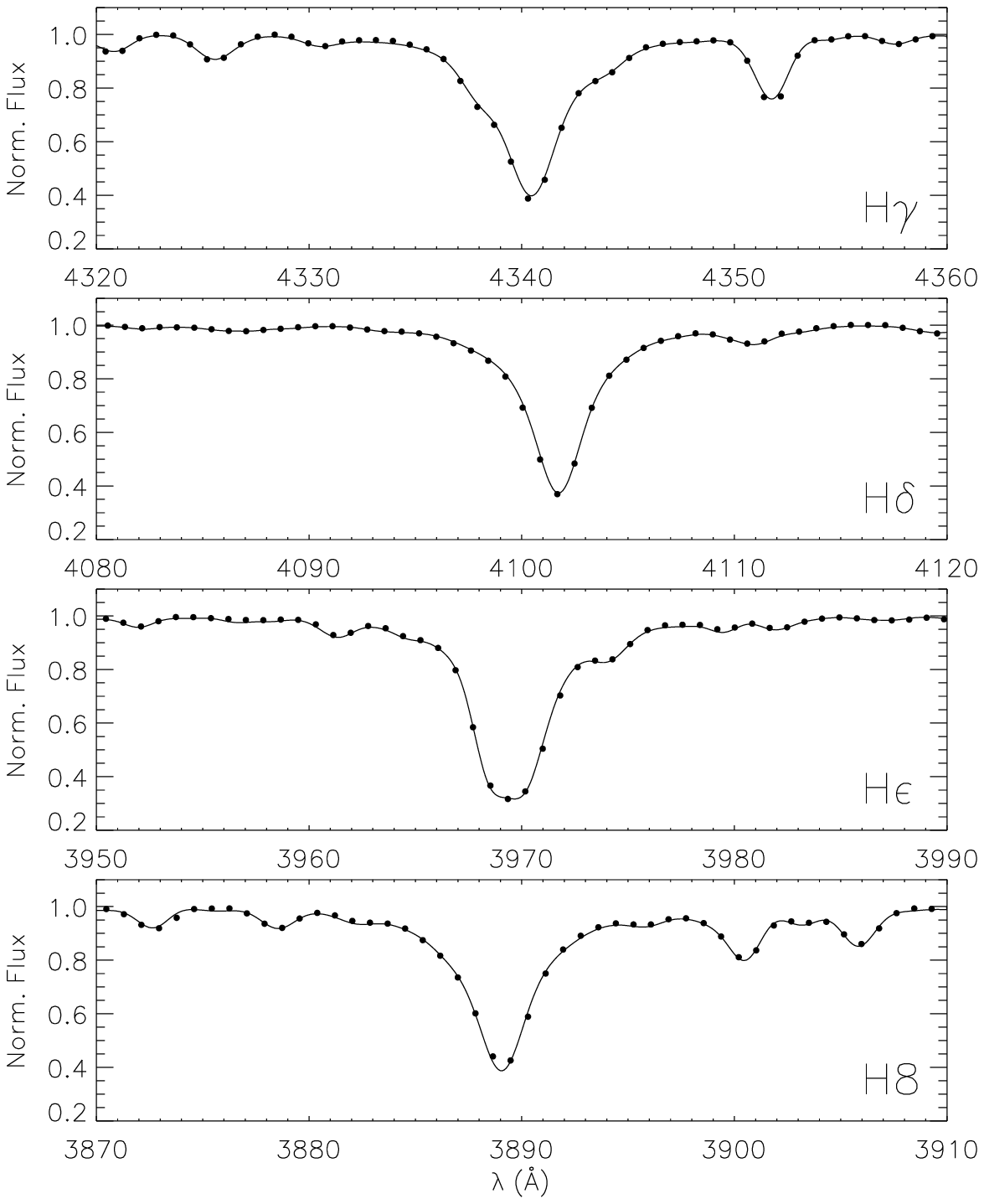}\\[-4mm]
\caption{Comparison of our {\sc Foces} spectrum (full line) with a
longslit spectrum from the NStars project \citep[dots]{gray03} 
in the region of the Balmer lines H$\gamma$ to H8 for Deneb (A2\,Ia) to 
assess the quality of the normalisation of our echelle spectra.
Note that the {\sc Foces} spectrum was was artificially degraded in
resolution to match the 1.8\,{\AA} resolution of the longslit data.}
\label{denebnormalisation}
\end{figure}

A semi-automatic pipeline was used for the reduction of the
{\sc Foces} data \citep{Pfeiffer}, performing subtraction of bias and dark current, 
flatfielding, wavelength calibration, rectification and merging of echelle orders. 
As the {\sc Foces} observations were obtained over several 
nights in different years, the conditions and quality varied for each set of exposures.
Before the data reduction procedure a median filter was 
applied to the raw data to remove cosmic ray hits without losing spectral information.
Special attention had to be paid to the merging of orders, because key 
features like the hydrogen lines extend over adjacent orders and their analysis 
is sensitive to line shape. Moreover, meticulous normalisation procedures were 
applied (within the reduction pipeline), since imprecise continuum 
determination may lead to incorrect abundance and parameter
determination. Comparison of our data with longslit spectra of fundamental MK
standards obtained within the
NStars project\footnote{\tt http://stellar.phys.appstate.edu/}
\citep{gray03} was possible for some cases and showed good agreement.
Figure~\ref{denebnormalisation} exemplifies such a comparison of 
a NStars project 1.8\,{\AA}-resolution spectrum around the Balmer lines
H$\gamma$ to H8 of Deneb (taken at the Dark Sky Observatory)
with our {\sc Foces} spectrum, which was artificially degraded in 
resolution to match that of the NStars project spectrum.
Finally, for several objects multiple exposures were taken and 
combined to enhance the quality and filter out artifacts.
 
Additional objects were observed in 2007 at the European Southern Observatory 
on La Silla, using {\sc Feros} \citep[Fiber-fed Extended Range Optical
Spectrograph,][]{FEROS} on the 2.2\,m telescope. The spectra cover the 
wavelength range from $\sim$3600 to 9200\,{\AA} at
$R$\,$\approx$\,48\,000. The raw spectra were mainly processed by 
the ESO automatic reduction pipeline. In addition to spectra of our
science targets, a template spectrum of a subdwarf B star was obtained and 
reduced with the pipeline. By dividing this spectrum through a well-fitting 
model we could successfully filter out all artifacts from the data 
reduction process, providing -- after proper smoothing to remove
localised noise relics -- a template function for the continuum
rectification of the science target spectra. This procedure provides
an objective and reproducible means for the normalisation process.

Additional high-quality spectra of bright BA-type supergiants were extracted
from the database of the UVES Paranal Ob\-ser\-vatory Project \citep{UVESPOP}. 
These data combine a high resolution of $R$\,$\approx$\,80\,000 and high S/N 
with large wavelength coverage from 3040 to 10\,400\,{\AA}, completing our sample. 

\begin{figure*}
\centering
\includegraphics[width=.97\linewidth]{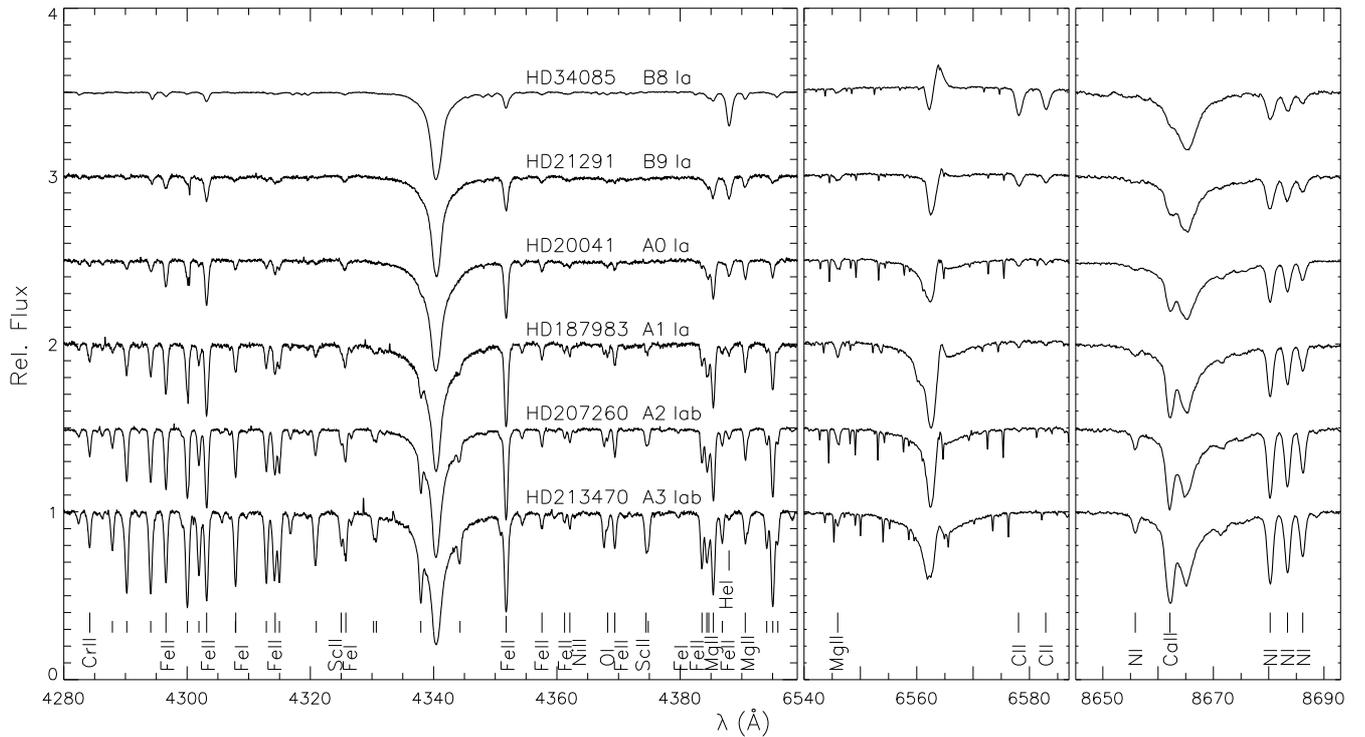}\\[-4mm]
\caption{Normalised spectra of an $T_{\rm eff}$-sequence of sample stars
around H$\gamma$ (left), H$\alpha$ (middle) and Pa13 (right panel), 
from the hot (top) to the cool
end of the parameter range investigated here (bottom). The major spectral features 
are identified, short vertical marks indicate \ion{Ti}{ii} lines.
The H$\alpha$ region is contaminated by narrow telluric lines.
Vertical shifts of the spectra by multiples of 0.5 have been applied for clarity.}
\label{spectra}
\end{figure*}

All together, this comprises the most comprehensive sample of
Galactic BA-type supergiants\footnote{Besides a supergiant nature
also a scenario as a post-AGB candidate is considered for HD\,195324 \citep{Szczerba07}
because of its infrared excess measured by IRAS \citep{Jaschek91,Oudmaijer92}. 
We will address the evolutionary state of this interesting object
based on surface abundance patterns in Paper\,II.}
with high-quality spectra considered for quantitative analysis to date.

The high-quality spectra allow a much closer look to be taken on 
trends in spectral line strengths and line ratios than possible with
traditional classification spectra at much lower resolution. We used
the opportunity to reassess the spectral classification of the sample
stars. Starting point of our approach were the anchor points of the MK
system (as identified in Table~\ref{observations1}), 
supplemented by MK primary standards as given by \citet{JoMo53}, see
also Table~\ref{observations1}. These cover about half of our sample
stars. Spectral types for the remaining stars were assessed on basis
of the helium and metal lines, luminosity classes by the width of the
Balmer lines. In general, good agreement with the classification as
obtained from SIMBAD was found. Maximum changes indicated by our
inspection, for about half of the non-MK stars, amount to one spectral
subtype or to one subtype (higher or lower) within the supergiant luminosity 
class. Our reclassification is indicated in Table~\ref{observations1}.
We note that two of the primary MK standards (HD\,14489 and HD\,223960) 
were found to differ significantly in spectral morphology from the
stars of similar original spectral type. Also the luminosity appears
overestimated for HD\,212593, which shows a nearly symmetric H$\alpha$
absorption profile. We therefore propose a (slight)
reclassification for these stars as well, based on the 
{\em available high-resolution spectra}.

Spectra of several sample stars are displayed in Fig.~\ref{spectra},
for three strategic wavelength regions around H$\gamma$, H$\alpha$ and
Pa13. A sequence in spectral type from B8 to A3 is shown, the earliest and 
latest types observed here. Note the rapid strengthening of the metal
lines towards lower effective temperatures, and their enormous
increase in number. The spectra were shifted to 
the laboratory rest frame and the major spectral features are
identified, many of which were analysed in our project.

Not unexpected are the asymmetric H$\alpha$ line profiles, indicating
the presence of a stellar wind of varying strength in the different
stars. P-Cygni-like profiles like in $\beta$\,Ori are an exception,
occurring only in the highest-luminosity objects (being most extreme
in HD\,12953). Typical for many stars of the sample is excess emission
in the red wing of H$\alpha$ like for the A0-A3 stars in
Fig.~\ref{spectra}. Only the luminosity class Ib stars show H$\alpha$
completely in absorption, with only small asymmetries being present.
See the atlas of \citet{Verdugo99a} for more examples on A-type
supergiant spectra.

In addition to the spectra, which constitute the principal data
for the analysis, various (spectro-)photometric data were adopted
from the literature for constructing spectral energy distributions. 
Johnson $UBV$-magnitudes were taken from \citet{mermilliod} 
(see Table \ref{observations1}), which are means of previously published 
photoelectric data, and $JHK$-magnitudes from the Two Micron All 
Sky Survey \citep[2MASS]{2MASS,Skrutskie06}. Additionally, 
flux-calibrated, low-dispersion 
spectra observed with the International Ultraviolet Explorer (IUE) were
extracted from the MAST archive\footnote{\tt http://archive.stsci.edu/},
where available (i.e.~for 16 objects). For seven objects only
high-resolution IUE spectra are present in the MAST archive, which are
nevertheless useful for our purposes as they were observed using a
large aperture (i.e.~flux-losses should have not occured).   
These data cover the range from 1150 to 
1980\,{\AA} for the short (SW) and from 1850 to 3290\,{\AA} for the long wavelength 
(LW) range camera. Typically, both wavelength ranges were observed the
same day. A summary of the individual spectra used in the present work
(data ID number and observation date) is given in Table~\ref{IUE}. 
Finally, photometric data in the Str\"omgren system was adopted from
the catalog of \citet[not tabulated here]{HauMer98}. 

\section{Quantitative analysis\label{sectanalysis}}  

\subsection{Models and analysis methodology}

Advanced models and robust analysis techniques are required in order 
to achieve the targeted accuracy and precision in stellar parameter and abundance determination. 
A suitable analysis methodology was presented by \citet{Norbert}, which we adopted 
for the present work, with some extensions as summarised below. 
The method is based on a hybrid non-LTE approach, in which non-LTE line-formation 
computations are performed on top of LTE model 
atmospheres\footnote{First results obtained with hydrodynamic, fully line-blanketed 
non-LTE atmospheres computed with {\sc Cmfgen} \citep{Chesneau10} and {\sc PoWR}
(W.-R. Hamann, priv. comm.) indicate that the photospheric spectra of
BA-type supergiants are described well by our approach.}.

\begin{table}[t]
\centering
 \setlength{\tabcolsep}{.18cm}
\caption{IUE spectra used in this study. \label{IUE}}
 \begin{tabular}{rlllll}
 \noalign{}
\hline\hline
\#  & Object    &  SW      & Date       & LW     & Date\\
\hline\\[-3mm]
1  & HD12301  &  P07282  & 01/12/1979 & R06276 & 01/12/1979\\
2  & HD12953  &  P42698  & 12/10/1991 & P21249 & 15/09/1991\\ 
3  & HD13476  &  \ldots  & \ldots     & P29471\tablefootmark{a} & 07/11/1994\\
4  & HD13744  &  \ldots  & \ldots     & P31590\tablefootmark{a} & 14/10/1995\\
5  & HD14433  &  \ldots  & \ldots     & P09166\tablefootmark{a} & 24/09/1986\\
6  & HD14489  &  P21812  & 19/12/1983 & \ldots & \ldots    \\
7  & HD20041  &  P56064  & 09/10/1995 & P31581 & 09/10/1995\\ 
8  & HD21291  &  P07280  & 01/12/1979 & R06274 & 01/12/1979\\
9  & HD39970  &  P56171  & 09/11/1995 & P31675 & 09/11/1995\\
10 & HD46300  &  P56165  & 08/11/1995 & P31667 & 08/11/1995\\
12 & HD187983 &  P48688  & 19/09/1993 & P26414 & 19/09/1993\\
13 & HD197345 &  P09133  & 26/05/1980 & R07864 & 26/05/1980\\
14 & HD202850 &  P15099  & 25/09/1981 & R11614 & 23/09/1981\\
15 & HD207260 &  P03368  & 17/11/1978 & R02957 & 17/11/1978\\
16 & HD207673 &  \ldots  & \ldots     & P31682\tablefootmark{a} & 10/11/1995\\
17 & HD208501 &  P55805  & 03/09/1995 & P31403 & 03/09/1995\\
18 & HD210221 &  P18682  & 28/11/1982 & R14745 & 28/11/1982\\
19 & HD212593 &  P33852  & 03/07/1988 & P13556 & 03/07/1988\\
20 & HD213470 &  \ldots  & \ldots     & P29702\tablefootmark{a} & 13/12/1994\\
22 & HD223960 &  \ldots  & \ldots     & R08994\tablefootmark{a} & 11/10/1980\\
24 & HD34085  &  P31880  & 18/09/1987 & P11654 & 18/09/1987\\
25 & HD87737  &  P08566  & 26/03/1980 & R07305 & 26/03/1980\\
33 & HD102878 &  \ldots  & \ldots     & P28097\tablefootmark{a} & 09/05/1994\\
\hline\\[-5mm]
\end{tabular}
\tablefoot{
\tablefoottext{a}{high-resolution spectrum}}
\end{table}

\begin{figure*}[ht!]
\centering
\includegraphics[width=.75\linewidth]{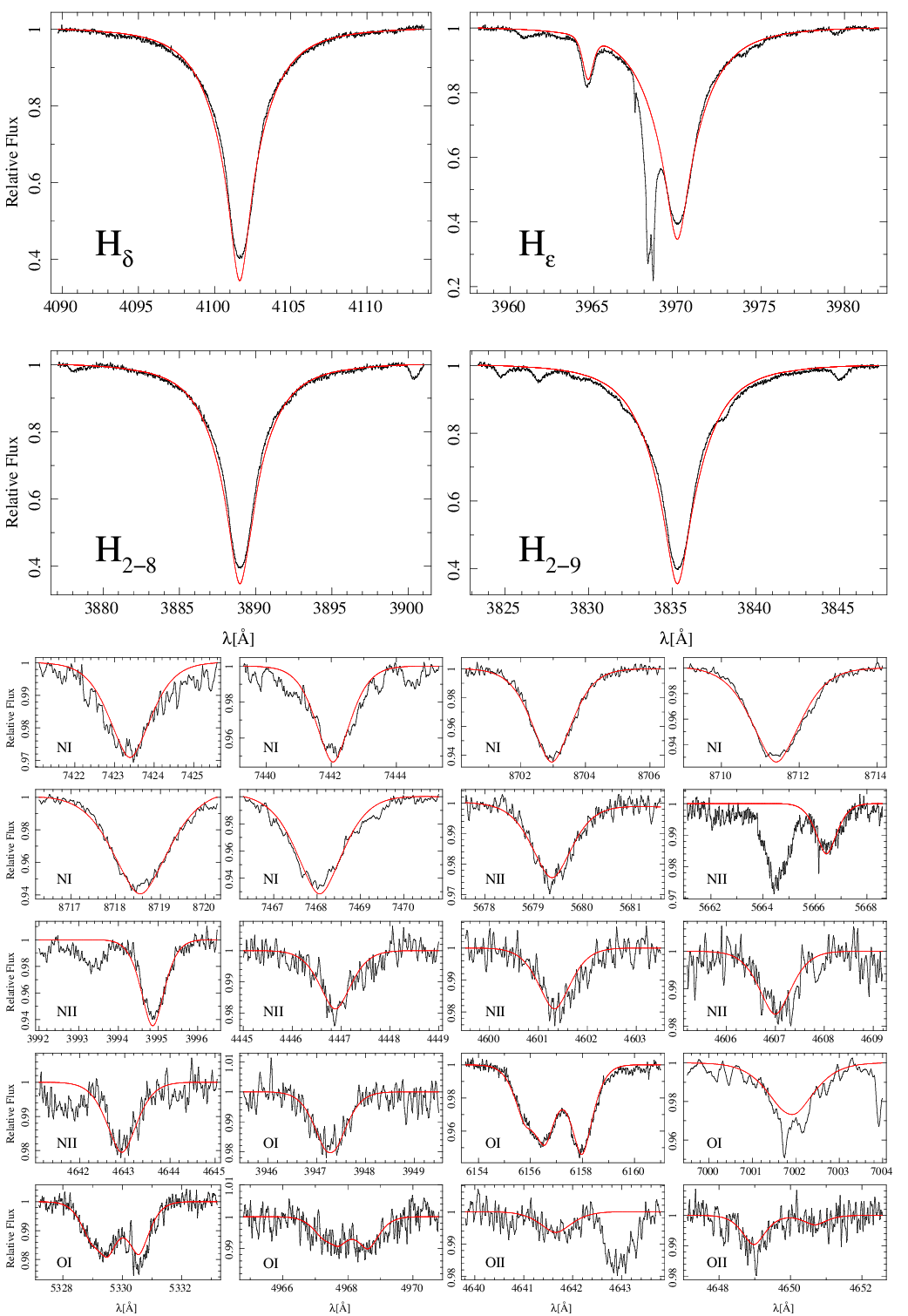}\\[-2mm]
\caption{Example for a set of diagnostic lines used in the determination of 
the atmospheric parameters $T_{\mathrm{eff}}$ and $\log g$ in the supergiant 
HD\,106068 (B8\,Iab), consisting of hydrogen, nitrogen and oxygen lines. 
The models calculated for our final parameters (red lines) are compared to 
observation (black lines). The same values of oxygen and nitrogen abundance 
are adopted for all lines, regardless of ionization stage. The ionization 
equilibria and the Balmer lines unaffected by the stellar wind are 
reproduced simultaneously. Note that various line blends of stellar, 
interstellar and terrestrial nature seen in the plots are 
exempted from the fitting process. }
\label{analysis}
\end{figure*}  

In brief, LTE model atmospheres were computed using the {\sc Atlas9} code 
\citep{Kurucz93b} 
with some additional modifications necessary to allow model convergence close to
the Eddington limit \citep{NorbertC}. Line blanketing was accounted for via 
Kurucz' Opacity Distribution Functions \citep{Kurucz93a}. Based on these model 
atmospheres the 
non-LTE calculations were performed with recent versions of {\sc Detail} 
and {\sc Surface} \citep[both updated by K. Butler]{Giddings,Butler}. The
coupled radiative transfer and statistical equilibrium equations were solved with 
{\sc Detail}, providing non-LTE level populations. Subsequently, synthetic
spectra were computed with {\sc Surface}, using refined line-broadening theories. 
State-of-the-art model atoms according to Table~\ref{atoms} 
were adopted for the calculation of the model grids. 

\begin{table}[t!]
\footnotesize
\caption[]{Model atoms for non-LTE calculations.\\[-6mm] \label{atoms}}
\setlength{\tabcolsep}{.15cm}
\begin{tabular}{ll}
\hline
\hline
\footnotesize
            Ion     &  Model atom \\
\hline\\[-3mm] 
     H          &  \citet{NorbertH}\\
\ion{He}{i}     &  \citet{NorbertHe}\\
\ion{C}{i/ii}   &  \citet{NorbertC,Nieva06,Nieva08}\\
\ion{N}{i/ii}   &  \citet{NorbertN}\\
\ion{O}{i/ii}   &  \citet{NorbertO,BeBu88}, updated\\
\ion{Mg}{i/ii}  &  \citet{NorbertMg}\\
\ion{S}{ii/iii} &  \citet{Vrancken96}, updated\\
\ion{Ti}{ii}    &  \citet{Becker98}\\
\ion{Fe}{ii}    &  \citet{Becker98}\\
\hline\\[-5mm]
\end{tabular}
\end{table}

Overall, around 25\,000 model spectra per element were combined into 5-dimensional
grids. The parameter space covered ranges from 8300\,K to 15\,500\,K in effective temperature
$T_{\mathrm{eff}}$ (in steps of 250-500\,K), from 2.50 in surface gravity 
$\log g$ (cgs units) to the convergence limit at 0.95 (lower
$T_{\mathrm{eff}}$-limit) to 1.90 (upper $T_{\mathrm{eff}}$-limit,
in steps of 0.1\,dex), from 3\,km\,s$^{-1}$
to 8\,km\,s$^{-1}$ in microturbulence $\xi$ (1\,km\,s$^{-1}$ steps) and 
from 0.09 to 0.15 in helium abundance (by number, steps of 0.015).
Finally, sufficient coverage in elemental abundance (typically
1\,dex in steps of 0.25\,dex) for the respective lines
investigated is provided.

Models of Doppler-shift profiles were computed by disk integration, using the 
methods described in \citet{gray}. We divided the apparent disk of a star in 16 
million parts of equal size. For each disk position, the radial-tangential 
velocity distribution, representing macroturbulence\footnote{A
physical explanation of the `macroturbulent' velocity parameter was only
recently suggested for hot stars, likely being the collective effect of
pulsations \citep{aerts09}, see also \citet{Sergio10}.}, was projected on the 
line of sight, assuming an equal amount of radial and tangential motion. 
The result was then Doppler-shifted according to stellar rotation. Finally, we 
applied a linear limb-darkening law to determine the weighting factors for 
integration. The resulting profiles are characterised by three parameters: 
the projected rotational velocity $v \sin i$, the macroturbulent
velocity $\zeta$, and the limb-darkening coefficient $\epsilon$.

Theoretical and observed line profiles were compared via the software package 
{\sc Spas} \citep[Spectral Plotting and Analysis Suite,][]{Hirsch}. It provides 
the means to interpolate between model grid points for up to three parameters 
simultaneously and allows to apply instrumental and macrobroadening functions 
to the resulting theoretical profiles. Furthermore, the program uses the downhill 
simplex algorithm to minimise $\chi^2$ in order to find a good fit to the 
observed spectrum. 

\begin{figure}[t]
\centering
\includegraphics[width=.995\linewidth]{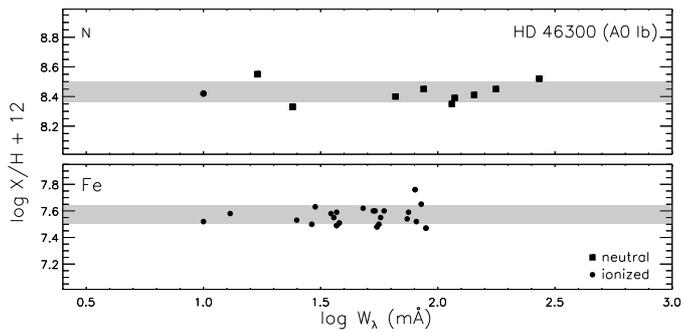}
\caption{Abundance values derived from single lines are presented 
versus respective equivalent widths for a selection of lines from
neutral and ionized nitrogen and iron 
in the spectrum of HD\,46300\,(A0\,Ib). The 1$\sigma$-scatter around the mean value is 
indicated by the gray bands.}
\label{equiw}
\end{figure}

\subsection{Spectroscopic indicators}
In order to find a globally satisfying solution, it was necessary to derive 
the basic atmospheric parameters $T_{\mathrm{eff}}$, $\log g$, $\xi$ and $y$ 
in an iterative process, utilising different spectral indicators on the way. 
Here we discuss the individual steps of this procedure.

\subsubsection{Effective temperature and surface gravity} 
The hydrogen lines and the ionization equilibria of the metals react 
sensitive to variations of effective temperature and surface gravity. 
This dependency was demonstrated for the Stark-broadened hydrogen
Balmer lines as well as for e.g. the Mg\,{\sc i/ii} lines by
\citet{Norbert} and \citet{Schiller}. While neither the hydrogen lines nor a 
single ionization equilibrium alone are sufficient to constrain $T_{\mathrm{eff}}$ 
and $\log g$ unambiguously due to a degeneracy in the solutions,
a combination of at least two of them is, for constant $\xi$ and $y$.  
Note that in the low-temperature regime of our sample, where N\,{\sc ii} lines are 
too weak to be analysed, the C\,{\sc i/ii} ionization equilibrium is
often available to confirm the results obtained from the Mg\,{\sc i/ii} lines, 
while in the high-temperature regime O\,{\sc i/ii} becomes available
in addition to N\,{\sc i/ii}. For the majority of the stars two or in some
cases even three metal ionization equilibria can thus be utilised for
the atmospheric parameter determination. H$\alpha$ (see
Fig.~\ref{spectra}) and, in the most
luminous objects, even H$\beta$ and H$\gamma$ can be affected by the
presence of a significant stellar wind. Consequently, these lines have to 
be omitted in such cases for the derivation of basic atmospheric parameters 
in our hydrostatic approach. H$\beta$ and H$\gamma$ are therefore
considered for {\em most} stars, and several of the higher Balmer lines
are analysed in {\em all} stars, with the exact number depending on the
wavelength coverage of the spectrograph, and the S/N reached in the
blue orders.
If ionization equilibria for more than one element are available 
for analysis, the results show typically high consistency. An example for the 
agreement of a final solution with observation is shown in Fig.~\ref{analysis}. 

\begin{figure*}
\centering
\includegraphics[width=.82\linewidth]{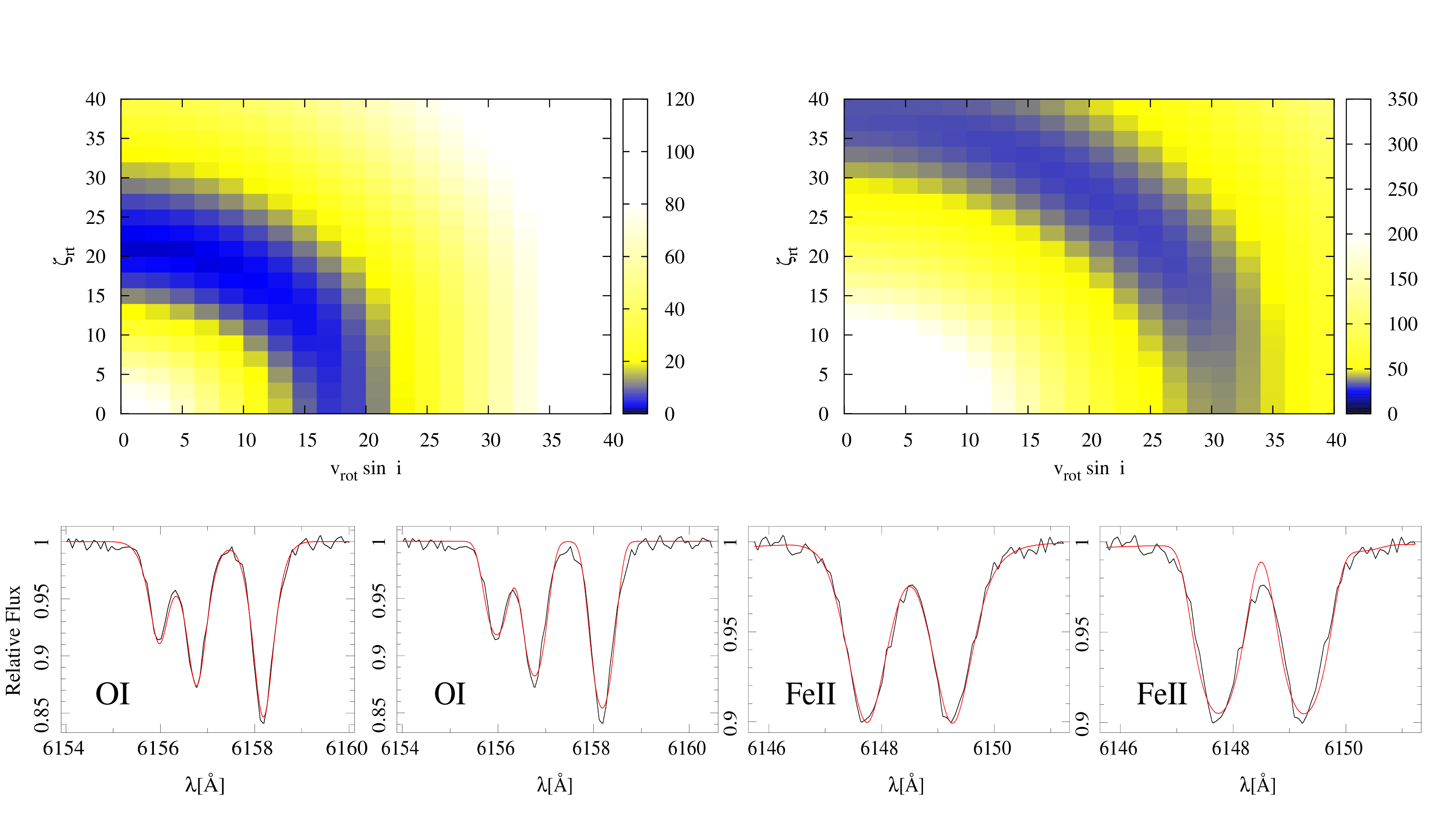}
\caption{Examples for the derivation of $v \sin i$ and $\zeta$ from 
the \ion{O}{i} triplet 
around 6157\,{\AA} (left) in HD\,207673 and the \ion{Fe}{ii} pair around 
6148\,{\AA} (right) in HD\,14433. The upper panels show contour plots of 
a goodness-of-fit parameter in ($v \sin i$, $\zeta$) space, lower values 
representing a better fit. The lower panels display the best line-profile-fits with 
and without macroturbulence (left and right, respectively).}
\label{vrotzeta}
\end{figure*}

\subsubsection{Microturbulence and helium abundance}
The microturbulent velocity is usually determined by demanding that the 
abundances indicated by the lines within an ion are independent of equivalent 
width. This is basically equivalent to finding the microturbulence
value that minimises the line-to-line scatter in abundance, which is our preferred 
method (as we use line-profile fits for this like for the rest of the
analysis). 
Within the framework of the present paper the nitrogen spectra are 
the most useful indicators for microturbulence as they provide
many lines of widely differing strength, while e.g.~the more
extensive \ion{Fe}{ii} line analysis (to be discussed in Paper\,III),
among others,
can be used to verify the $\xi$-determination. An example 
is shown in Fig.~\ref{equiw}, using the traditional form of 
illustration.
Equivalent widths were determined using direct integration over
the line profile. Linear regression curves to the data
points are consistent with a slope of zero in both cases.
A single value for $\xi$ was found to be consistent for all 
elements analysed in non-LTE in the individual stars, see 
Fig.~13 of \citet{Norbert} for a comprehensive
discussion in the case of HD\,87737. Note that while the Balmer-lines are 
virtually insensitive to microturbulence variations, this quantity 
influences the parameter determination based on ionization equilibria.

A change in the atmospheric helium abundance gives rise to a modified mean 
molecular weight of the atmospheric plasma, which has similar effects on 
model predictions than a variation of $\log g$
\citep{Kudritzki73,Norbert}. Therefore, the helium abundance $y$ has to be 
constrained simultaneously with the other atmospheric parameters. 
This is done via line-profile fitting of helium lines. Results for the individual 
\ion{He}{i} line abundances in the sample stars can be found in Paper\,II.

In practice, estimates for $\xi$ and $y$ are used to determine initial values 
for $T_{\mathrm{eff}}$ and $\log g$ in a first iteration step (using
elemental abundances as third fit parameter), which are in turn 
adopted to derive $\xi$ and $y$ in a next step. Convergence is quickly
achieved using our comprehensive model grids and {\sc Spas}.

\subsubsection{Projected rotational velocity and
macroturbulence\label{rotzeta}}
In order to perform line-profile-fitting it is necessary that the models 
are able to reproduce the line-shapes accurately. To achieve this our model 
spectra are convolved with the functions for rotational and macroturbulent 
broadening computed as described earlier. A value of 0.5 for the limb-darkening 
coefficient $\epsilon$ was adopted throughout. Tests showed little sensitivity for this 
parameter within acceptable bounds, which is only slowly varying around our
canonical value throughout the atmospheric parameter and spectral wavelength
range investigated here \citep[see e.g.][]{Wade}. 

We chose the same set of highly suitable metal lines to derive $v \sin
i$ and $\zeta$ in most of our targets, including the \ion{Mg}{ii} line at 
4390\,{\AA}, the \ion{Mg}{i} line at 5183\,{\AA}, the \ion{Fe}{ii}
pair around 6148\,{\AA}, and the \ion{O}{i} triplet around 6157\,{\AA}. 
However, for the two hottest objects in our sample only the \ion{S}{ii} line at 
5354\,{\AA} was used. Examples from the fitting are shown in
the upper panels of Fig.~\ref{vrotzeta}, displaying a goodness-of-fit 
parameter in ($v \sin i$, $\zeta$) space. While in theory this should allow us
to identify the optimum values for $v \sin i$ and $\zeta$, we found
disentanglement of the effects from rotational velocity and 
macroturbulence to be difficult because of the rather smeared
contours. This is similar to other recent studies of 
rotational broadening in BA-type supergiants \citep{Verdugo,Ryans}.
Values of $v \sin i$ and $\zeta$ representing the best fits show a scatter 
from line to line because of this degeneracy in the solutions. 
Fits without macroturbulence contribution agree very well in the 
resulting $v \sin i$ for different absorption features. However, 
they can not reproduce observations 
in such a satisfactory way like fits that consider both effects, see
the lower panels of Fig.~\ref{vrotzeta} for a comparison.
Overall, the uncertainties of $v\sin i$ and $\zeta$ are typically about
$\pm$5\,km\,s$^{-1}$ because of the degeneracy of the solutions, and
about $\pm$3\,km\,s$^{-1}$ for the stars with the sharpest lines.

\subsubsection{Abundances and metallicity}
Abundances for all ionization stages involved in the atmospheric parameter 
determination were derived, as summarised in Table~\ref{parameters}. For
this, the individual lines of an ion were fitted and the mean of the resulting 
abundances was taken. For the sake of brevity we postpone a presentation of 
individual line abundances for the sample stars to  
Papers\,II and III, where the complete data will be discussed, irrespective of the
availability of ionization equilibria.

Overall, good to excellent agreement between different ionization 
stages of the elements was found, even in cases where more than one 
ionization equilibrium was evaluated, indicating that major bias is
likely absent in our atmospheric parameter determination. The statistical 
uncertainties arising 
from noise prove to be small due to the high quality of our spectra. Only for 
the weakest lines used in our analysis they are comparable to the line-to-line 
scatter. Therefore we chose the 1$\sigma$ standard deviation based on
the individual line abundances as a conservative error estimate for
the ion abundances. In this approach it is difficult to assign 
an uncertainty value for an ion where only one line is available for analysis. 
An inspection of the data in Table~\ref{parameters} indicates that a
characteristic estimate in such a case may be 0.10\,dex.

Low star-to-star scatter in metallicity was found in recent analyses of 
early B-type dwarfs -- the progenitors of BA-type supergiants on the
main sequence -- in the solar neighbourhood 
\citep{Standard,Nieva11,Nieva12}, using similar analysis techniques. 
Given the low sensitivity of 
our analysis to small changes in metallicity \citep{Norbert}, this parameter was held fixed 
throughout our model grid to slightly subsolar values 
\citep[with respect to the solar standard of][]{Grevesse} in
accordance with the B-dwarf results (the cosmic abundance standard).
This approximation was proved correct {\em a-posteriori} by our abundance analysis for
most of the sample stars. We reanalysed the most metal-rich stars of our sample 
later on by means of fine-tuned micro-grids, adopting a higher value for the metallicity
in order to rule out any possible bias due to this.

\begin{figure*}[t]
\centering
\includegraphics[width=.43\linewidth]{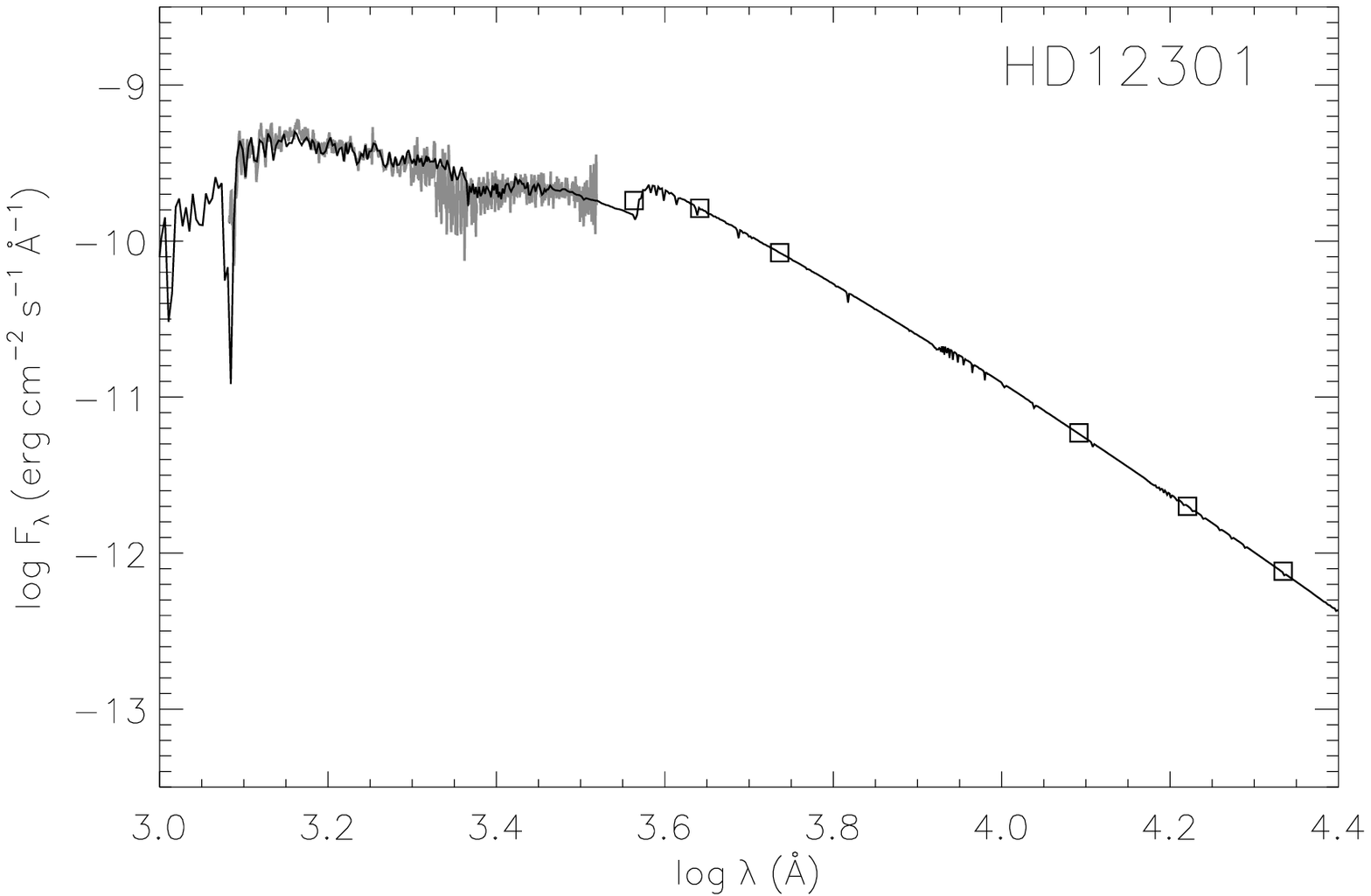}
\includegraphics[width=.43\linewidth]{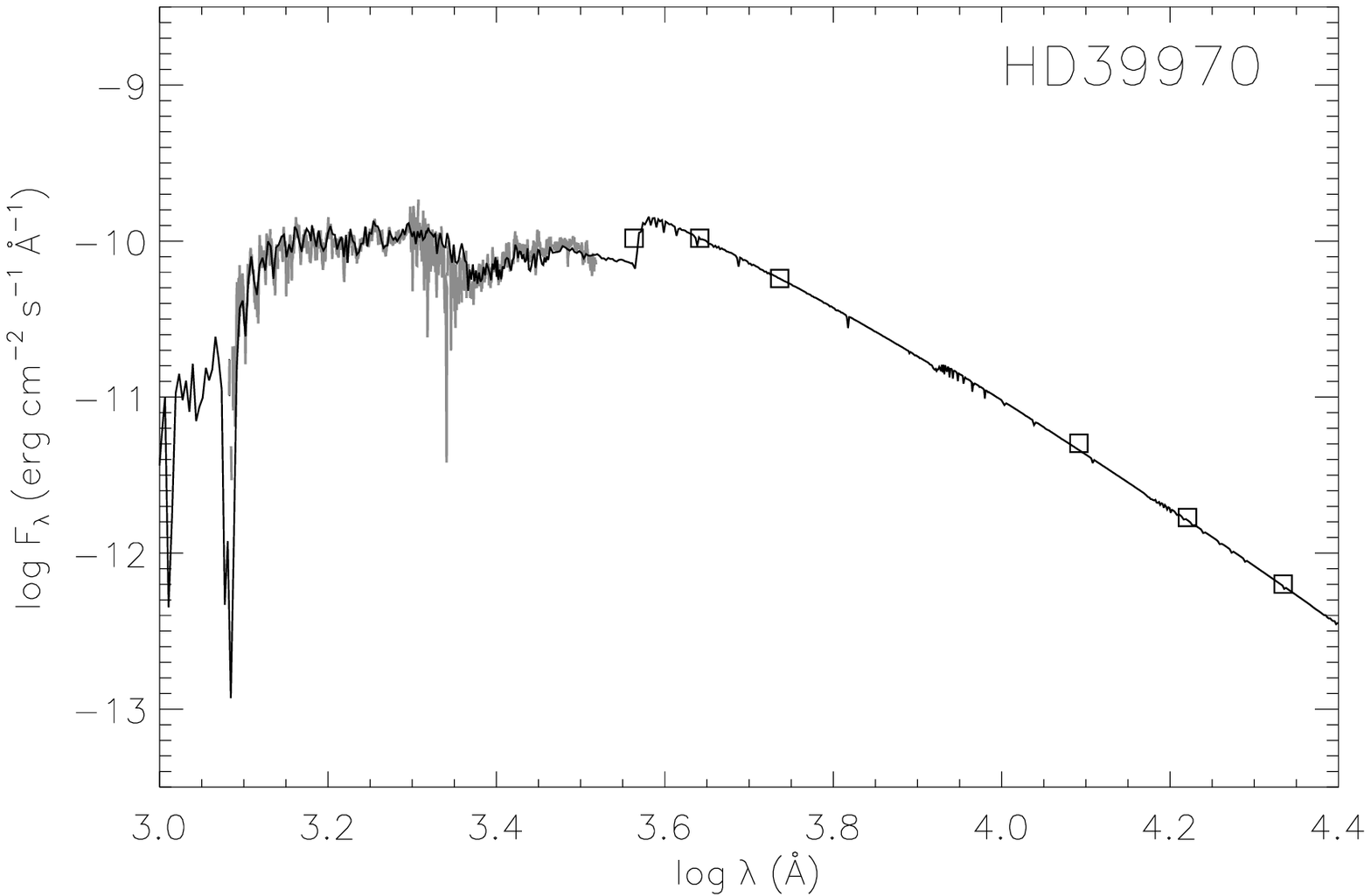}
\caption{Examples of comparisons of {\sc Atlas9} model fluxes (black lines)
with UV-spectrophotometry from the IUE satellite (gray lines)
and with photometric measurements -- $UBV$ from Mermilliod \&
Mermilliod~(1994), $JHK$ from 2MASS --
for two of our sample stars. The SEDs shown are dereddened according
to the values of $E(B-V)$ and $R_V$ in Table~\ref{parameters}, and normalised in $V$.}
\label{SED}
\end{figure*}

\subsubsection{Spectral energy distribution\label{SEDsection}}
As a verification of our spectroscopic analysis we compared the {\sc Atlas9} 
fluxes computed for our final parameters with photometric data in the
optical and near-IR, and UV spectrophotometry. Thus, we investigated
whether the models reproduce also the spectral energy distributions
(SEDs) of the stars, and as a consequence also their global energy output.

We transformed the $UBV$ and $JHK$ magnitudes into absolute fluxes using zeropoints 
from \citet{Bessell} for Johnson photometry and from \citet{cohen} for the 
2MASS photometric system. Note that we lowered the $U$ photometric zero point by 3\%
based on discrepancies between theory and observation that were 
found by \citet{vegaSED} in the respective part of the UV-flux for the photometric
standard~star~Vega.

The observed fluxes needed to be dereddened for the comparison with
the model fluxes. For this, we adopted the mean interstellar extinction law of
\citet{Cardelli}, which is depending on one parameter, the ratio of
total-to-selective extinction $R_V$\,$=$\,$A_V/E(B-V)$. The colour
excess $E(B-V)$ was determined per star as the difference between the
theoretical colour computed from the {\sc Atlas9} flux and the observed 
colour. Then, $R_V$ was varied until a good overall match of the
theoretical and the observed SED was achieved, see Fig.~\ref{SED}
for examples of final results. The typical uncertainty in $R_V$ is 
$\pm$0.2.
This could be done also for the twelve
stars without IUE spectrophotometry, though lacking the very thorough
constraints by the UV fluxes.

In general, as good agreement between the model and the observed SED 
is found as exemplified in Fig.~\ref{SED}. Notable
deviations occur only in three cases. An IR excess is found for HD\,21291, 
while the UV-flux is in excellent agreement. We attribute this to HD\,21291 
being part of a close visual binary separated by only $2.39\arcsec$ 
\citep{HD21291}, which could not be resolved by 2MASS. An UV excess in
HD\,187983 can be explained by adding flux from an early B-type 
main-sequence companion, 
which is consistent with a radial-velocity variation found by 
\citet{HD187983}. There remain some problems in
reproducing the SED of HD\,12953, the most luminous object in our
sample, similar to those found for \object{HD\,92207} (A0\,Iae) by
\citet{Norbert}. Further investigations are required to address the
issue, but these are beyond the scope of the present~work.

Finally, a by-product of our investigations are bolometric corrections $B.C.$
for the sample stars. They were derived from the fluxes
of our final models, see Sect.~\ref{bolcorrections} for a discussion.

\section{Results}
\subsection{Atmospheric parameters of the sample stars}
Our results on the atmospheric parameters and some derived properties of 
the sample stars are summarised in Table~\ref{parameters}. 
For each object our internal sequential number and its HD-designation
are given, followed by our determinations of effective temperature,
surface gravity, microturbulent, projected rotational and macroturbulent 
velocity, colour excess, ratio of total-to-selective extinction and
bolometric correction. Finally, surface abundances of helium and -- 
for the cases where the respective ionization equilibria were employed 
for the atmospheric parameter determination -- abundances of metal
ions are summarised. The error margins denote 1$\sigma$-uncertainties
(an estimate of 0.10\,dex may be applied for ion abundances where 
only one line could be analysed, i.e.~those data without uncertainty
indicator). Uncertainies in $v\sin i$ and $\zeta$ amount typically to
$\pm$5\,km\,s$^{-1}$ ($\pm$3\,km\,s$^{-1}$ for the stars with the
sharpest lines) because of the degeneracy of the solution in the 
fitting of $v \sin i$ and $\zeta$, see Sect.~\ref{rotzeta}, 
and $\pm$0.2 in $R_V$, see Sect.~\ref{SEDsection}. The uncertainties
of the bolometric corrections vary from $\sim$0.02\,mag for the
coolest sample stars to $\sim$0.05\,mag for the hottest stars, see
Sect.~\ref{bolcorrections} for a more detailed discussion. 
This is the most comprehensive collection of such data from a
homogeneous non-LTE analysis of Galactic BA-type supergiants to date.
As expected, a continuous distribution over the temperature range from
8400 to 12\,700\,K 
is obtained, displaying no signs of systematic shifts or gaps stemming from the 
use of different temperature indicators.  

Tests with a bootstrapping method implemented in {\sc Spas} indicate 
formal uncertainties of less than 0.05\,dex in $\log g$ and 1\% in $T_{\mathrm{eff}}$, 
when all other parameters are held constant at the values of our final solution. 
This applies even for the lowest-quality spectra of our sample (at
S/N-ratio of about 120). Fits to 
individual Balmer lines -- the main indicators for $\log g$ throughout
the sample -- show a scatter of less than 0.05\,dex around the mean value in 
this parameter in the individual stars. This suggests that {\em major} 
systematic effects from order merging and normalisation in the hydrogen lines
may be absent. 

The line-to-line scatter in abundances derived from individual metal lines, 
however, indicates low-level systematic uncertainties from within our model atoms 
and the normalisation of weak lines. Small selection effects may also emerge in cases 
where only a few lines in one ion of the two analysed for establishing ionization balance
are available. Therefore, when line sets of sufficient quality for two or 
more elements were available, we compared different solutions using only one 
of the available ionization equilibria and the Balmer lines. While still showing
good agreement, the results imply slightly higher uncertainties than discussed earlier. 
Hence we chose our uncertainty estimates carefully across the
parameter space, remaining conservative in the values given in Table~\ref{parameters}. 
The largest discrepancies were found for the case when the last
residual \ion{Mg}{i} lines (\ion{Mg}{i}\,$\lambda$5183\,{\AA} being
the strongest) are about to vanish. In consequence, the
largest uncertainties occur in the transition region when 
N\,{\sc i/ii} starts to replace the Mg\,{\sc i/ii}-balance.
Overall, the typical conservative uncertainties in $T_{\mathrm{eff}}$ and 
$\log g$ are therefore $\sim$2\% (150--200\,K) and 
0.10\,dex (i.e.~$\sim$25\% in gravity), respectively,
accounting for both statistical and systematic errors. 
\begin{landscape}
\setlength{\tabcolsep}{.08cm}
\begin{table}
\centering
\caption{Stellar parameters of the program stars.}
 \label{parameters}
 \begin{tabular}{rlr@{$\pm$}lr@{$\pm$}lr@{$\pm$}lccr@{$\pm$}llr@{\hspace{3mm}}r@{$\pm$}lr@{$\pm$}lr@{$\pm$}lr@{$\pm$}lr@{$\pm$}lr@{$\pm$}lr@{$\pm$}lr@{$\pm$}lr@{$\pm$}l}
 \noalign{}
\hline\hline
\#  &Object   &\multicolumn{2}{c}{$T_\mathrm{eff}$}&\multicolumn{2}{c}{$\log\,g$}&\multicolumn{2}{c}{$\xi$}       & $v\sin\,i$\tablefootmark{a}  & $\zeta$\tablefootmark{a}   &\multicolumn{2}{c}{$E(B-V)$} & $R_V$\tablefootmark{b} &$B.C.$\tablefootmark{c}& \multicolumn{18}{c}{$\log X/{\rm H}$+12\tablefootmark{d}}\\[-.3mm]
\cline{15-32}
  &         &\multicolumn{2}{c}{K}               &\multicolumn{2}{c}{cgs}      &\multicolumn{2}{c}{km\,s$^{-1}$}&  km\,s$^{-1}$ & km\,s$^{-1}$ & \multicolumn{2}{c}{mag}     &       &  mag & \multicolumn{2}{c}{\ion{He}{i}}  & \multicolumn{2}{c}{\ion{C}{i}}&\multicolumn{2}{c}{\ion{C}{ii}}&\multicolumn{2}{c}{\ion{N}{i}}&\multicolumn{2}{c}{\ion{N}{ii}}&\multicolumn{2}{c}{\ion{O}{i}}&\multicolumn{2}{c}{\ion{O}{ii}}&\multicolumn{2}{c}{\ion{Mg}{i}}&\multicolumn{2}{c}{\ion{Mg}{ii}} \\[-.2mm]
\hline\\[-2mm]								       
1  & HD12301    & 12600  & 200  & 2.15  &   0.10  & 7  &   1  & 10  & 20  & 0.48  & 0.02   & 3.1 &  $-$0.72 & 11.06  &   0.06   & \multicolumn{2}{c}{    } & \multicolumn{2}{c}{    } & 8.16 & 0.05	       & 8.12 & 0.04		  & 8.72 & 0.06 	     & 8.65 & 0.03		& \multicolumn{2}{c}{	 } & \multicolumn{2}{c}{    } \\
2  & HD12953    & 9200   & 200  & 1.15  &   0.10  & 8  &   1  & 22  & 32  & 0.58  & 0.02   & 2.9 &  $-$0.13 & 11.12  &   0.03   & \multicolumn{2}{c}{    } & \multicolumn{2}{c}{    } & 8.40 & 0.03	       & \multicolumn{2}{c}{8.49} & \multicolumn{2}{c}{    } & \multicolumn{2}{c}{    } & \multicolumn{2}{c}{	 } & \multicolumn{2}{c}{    } \\
3  & HD13476    & 8500   & 150  & 1.40  &   0.10  & 6  &   1  & 12  & 24  & 0.59  & 0.02   & 2.8 &     0.06 & 11.16  &   0.08   & 8.18 & 0.12		 & 8.20 & 0.13  	    & \multicolumn{2}{c}{    } & \multicolumn{2}{c}{	} & \multicolumn{2}{c}{    } & \multicolumn{2}{c}{    } & 7.44 & 0.03		   & 7.46 & 0.04 \\
4  & HD13744    & 9500   & 250  & 1.55  &   0.15  & 6  &   1  & 12  & 29  & 0.76  & 0.02   & 3.1 &  $-$0.15 & 11.16  &   0.05   & \multicolumn{2}{c}{    } & \multicolumn{2}{c}{    } & \multicolumn{2}{c}{    } & \multicolumn{2}{c}{	} & \multicolumn{2}{c}{    } & \multicolumn{2}{c}{    } & 7.49 & 0.04		   & 7.39 & 0.07 \\
5  & HD14433    & 9150   & 150  & 1.40  &   0.10  & 7  &   1  & 17  & 29  & 0.56  & 0.02   & 3.1 &  $-$0.07 & 11.10  &   0.08   & \multicolumn{2}{c}{8.25} & 8.23 & 0.05  	    & \multicolumn{2}{c}{    } & \multicolumn{2}{c}{	} & \multicolumn{2}{c}{    } & \multicolumn{2}{c}{    } & 7.49 & 0.06		   & 7.40 & 0.04 \\
6  & HD14489    & 9350   & 250  & 1.45  &   0.15  & 7  &   1  & 13  & 35  & 0.38  & 0.02   & 3.1 &  $-$0.14 & 11.15  &   0.04   & \multicolumn{2}{c}{    } & \multicolumn{2}{c}{    } & 8.51 & 0.05	       & 8.54 & 0.08		  & \multicolumn{2}{c}{    } & \multicolumn{2}{c}{    } & \multicolumn{2}{c}{	 } & \multicolumn{2}{c}{    } \\
7  & HD20041    & 10000  & 200  & 1.65  &   0.10  & 7  &   1  & 14  & 37  & 0.75  & 0.03   & 3.1 &  $-$0.22 & 11.08  &   0.04   & \multicolumn{2}{c}{    } & \multicolumn{2}{c}{    } & 8.25 & 0.04	       & 8.35 & 0.04		  & \multicolumn{2}{c}{    } & \multicolumn{2}{c}{    } & \multicolumn{2}{c}{	 } & \multicolumn{2}{c}{    } \\
8  & HD21291    & 10800  & 200  & 1.65  &   0.10  & 7  &   1  & 32  & 33  & 0.46  & 0.02   & 3.1 &  $-$0.39 & 11.11  &   0.04   & \multicolumn{2}{c}{    } & \multicolumn{2}{c}{    } & 8.44 & 0.04	       & 8.48 & 0.03		  & \multicolumn{2}{c}{    } & \multicolumn{2}{c}{    } & \multicolumn{2}{c}{	 } & \multicolumn{2}{c}{    } \\
9  & HD39970    & 10300  & 200  & 1.70  &   0.10  & 7  &   1  & 2   & 45  & 0.43  & 0.02   & 3.5 &  $-$0.28 & 11.08  &   0.05   & \multicolumn{2}{c}{    } & \multicolumn{2}{c}{    } & 8.13 & 0.05	       & 8.22 & 0.14		  & \multicolumn{2}{c}{    } & \multicolumn{2}{c}{    } & \multicolumn{2}{c}{	 } & \multicolumn{2}{c}{    } \\
10 & HD46300    & 10000  & 200  & 2.15  &   0.10  & 3  &   1  & 0   & 14  & 0.07  & 0.02   & 2.5 &  $-$0.22 & 11.11  &   0.06   & \multicolumn{2}{c}{    } & \multicolumn{2}{c}{    } & 8.43 & 0.07	       & \multicolumn{2}{c}{8.42} & \multicolumn{2}{c}{    } & \multicolumn{2}{c}{    } & \multicolumn{2}{c}{	 } & \multicolumn{2}{c}{    } \\
11 & HD186745   & 12500  & 200  & 1.80  &   0.10  & 8  &   1  & 22  & 40  & 1.01  & 0.02   & 2.9 &  $-$0.72 & 11.06  &   0.05   & \multicolumn{2}{c}{    } & \multicolumn{2}{c}{    } & 8.37 & 0.05	       & 8.33 & 0.03		  & 8.77 & 0.01 	     & \multicolumn{2}{c}{8.77} & \multicolumn{2}{c}{	 } & \multicolumn{2}{c}{    } \\
12 & HD187983   & 9300   & 250  & 1.60  &   0.15  & 7  &   1  & 15  & 29  & 0.70  & 0.02   & 3.0 &  $-$0.08 & 11.08  &   0.07   & \multicolumn{2}{c}{8.25} & 8.31 & 0.12  	    & \multicolumn{2}{c}{    } & \multicolumn{2}{c}{	} & \multicolumn{2}{c}{    } & \multicolumn{2}{c}{    } & 7.58 & 0.04		   & 7.50 & 0.03 \\
13 & HD197345   & 8700   & 150  & 1.20  &   0.10  & 8  &   1  & 10  & 29  & 0.06  & 0.02   & 3.1 &    0.02  & 11.19  &   0.06   & 8.12 & 0.06		 & 8.07 & 0.09  	    & \multicolumn{2}{c}{    } & \multicolumn{2}{c}{	} & \multicolumn{2}{c}{    } & \multicolumn{2}{c}{    } & 7.50 & 0.05		   & 7.47 & 0.05 \\
14 & HD202850   & 10800  & 200  & 1.85  &   0.10  & 6  &   1  & 14  & 35  & 0.19  & 0.02   & 3.1 &  $-$0.39 & 11.20  &   0.07   & \multicolumn{2}{c}{    } & \multicolumn{2}{c}{    } & 8.68 & 0.04	       & 8.74 & 0.07		  & \multicolumn{2}{c}{    } & \multicolumn{2}{c}{    } & \multicolumn{2}{c}{	 } & \multicolumn{2}{c}{    } \\
15 & HD207260   & 8800   & 150  & 1.35  &   0.10  & 7  &   1  & 15  & 25  & 0.51  & 0.02   & 2.5 &    0.01  & 11.17  &   0.04   & 8.26 & 0.07		 & 8.17 & 0.09  	    & \multicolumn{2}{c}{    } & \multicolumn{2}{c}{	} & \multicolumn{2}{c}{    } & \multicolumn{2}{c}{    } & 7.51 & 0.05		   & 7.48 & 0.02 \\
16 & HD207673   & 9250   & 100  & 1.80  &   0.10  & 5  &   1  & 1   & 23  & 0.44  & 0.02   & 3.1 &  $-$0.08 & 11.11  &   0.10   & \multicolumn{2}{c}{8.18} & 8.16 & 0.11  	    & 8.48 & 0.03	       & \multicolumn{2}{c}{8.45} & \multicolumn{2}{c}{    } & \multicolumn{2}{c}{    } & 7.51 & 0.07		   & 7.44 & 0.04 \\
17 & HD208501   & 12700  & 200  & 1.85  &   0.10  & 8  &   1  & 16  & 56  & 0.82  & 0.02   & 2.7 &  $-$0.76 & 11.07  &   0.06   & \multicolumn{2}{c}{    } & \multicolumn{2}{c}{    } & 8.25 & 0.08	       & 8.22 & 0.07		  & \multicolumn{2}{c}{8.74} 		   & 8.77 & 0.01		& \multicolumn{2}{c}{	 } & \multicolumn{2}{c}{    } \\
18 & HD210221   & 8400   & 150  & 1.40  &   0.10  & 6  &   1  & 0   & 27  & 0.40  & 0.02   & 2.9 &    0.08  & 11.13  &   0.04   & 8.24 & 0.07		 & 8.17 & 0.01  	    & \multicolumn{2}{c}{    } & \multicolumn{2}{c}{	} & \multicolumn{2}{c}{    } & \multicolumn{2}{c}{    } & 7.51 & 0.06		   & 7.48 & 0.03 \\
19 & HD212593   & 11200  & 200  & 2.10  &   0.10  & 5  &   1  & 6   & 24  & 0.17  & 0.02   & 2.8 &  $-$0.46 & 11.15  &   0.06   & \multicolumn{2}{c}{    } & \multicolumn{2}{c}{    } & 8.45 & 0.02	       & 8.43 & 0.09		  & \multicolumn{2}{c}{    } & \multicolumn{2}{c}{    } & \multicolumn{2}{c}{	 } & \multicolumn{2}{c}{    } \\
20 & HD213470   & 8400   & 150  & 1.30  &   0.10  & 7  &   1  & 13  & 27  & 0.54  & 0.03   & 3.1 &    0.10  & 11.08  &   0.09   & \multicolumn{2}{c}{    } & \multicolumn{2}{c}{    } & \multicolumn{2}{c}{    } & \multicolumn{2}{c}{	} & \multicolumn{2}{c}{    } & \multicolumn{2}{c}{    } & 7.46 & 0.03		   & 7.50 & 0.06 \\
21 & BD+602582  & 11900  & 200  & 1.85  &   0.10  & 7  &   1  & 35  & 14  & 0.85  & 0.02   & 3.4 &  $-$0.58 & 11.14  &   0.07   & \multicolumn{2}{c}{    } & \multicolumn{2}{c}{    } & 8.51 & 0.04	       & 8.56 & 0.07		  & \multicolumn{2}{c}{    } & \multicolumn{2}{c}{    } & \multicolumn{2}{c}{	 } & \multicolumn{2}{c}{    } \\
22 & HD223960   & 10700  & 200  & 1.60  &   0.10  & 8  &   1  & 25  & 37  & 0.76  & 0.02   & 3.2 &  $-$0.37 & 11.12  &   0.08   & \multicolumn{2}{c}{    } & \multicolumn{2}{c}{    } & 8.57 & 0.07	       & \multicolumn{2}{c}{8.52} & \multicolumn{2}{c}{    } & \multicolumn{2}{c}{    } & \multicolumn{2}{c}{	 } & \multicolumn{2}{c}{    } \\
23 & HD195324   & 9200   & 150  & 1.85  &   0.10  & 4  &   1  & 3   & 20  & 0.56  & 0.02   & 2.6 &  $-$0.07 & 11.19  &   0.06   & 8.11 & 0.14		 & 8.08 & 0.06  	    & \multicolumn{2}{c}{    } & \multicolumn{2}{c}{	} & \multicolumn{2}{c}{    } & \multicolumn{2}{c}{    } & 7.59 & 0.03		   & 7.54 & 0.03 \\
24 & HD34085    & 12100  & 150  & 1.75  &   0.10  & 8  &   1  & 25  & 31  & 0.05  & 0.02   & 3.1 &  $-$0.65 & 11.08  &   0.05   & \multicolumn{2}{c}{    } & \multicolumn{2}{c}{    } & 8.50 & 0.04	       & 8.42 & 0.05		  & 8.73 & 0.04 	     & 8.76 & 0.06		& \multicolumn{2}{c}{	 } & \multicolumn{2}{c}{    } \\
25 & HD87737    & 9600   & 200  & 2.05  &   0.10  & 4  &   1  & 2   & 17  & 0.02  & 0.02   & 3.1 &  $-$0.14 & 11.17  &   0.05   & 8.23 & 0.10		 & 8.26 & 0.04  	    & 8.54 & 0.07	       & 8.49 & 0.06		  & \multicolumn{2}{c}{    } & \multicolumn{2}{c}{    } & 7.54 & 0.05		   & 7.52 & 0.04 \\
26 & HD91533    & 9100   & 150  & 1.50  &   0.10  & 6  &   1  & 20  & 29  & 0.33  & 0.02   & 3.5 &  $-$0.07 & 11.15  &   0.04   & \multicolumn{2}{c}{8.24} & 8.15 & 0.03  	    & \multicolumn{2}{c}{    } & \multicolumn{2}{c}{	} & \multicolumn{2}{c}{    } & \multicolumn{2}{c}{    } & 7.49 & 0.12		   & 7.46 & 0.08 \\
27 & HD111613   & 9150   & 150  & 1.45  &   0.10  & 6  &   1  & 17  & 27  & 0.39  & 0.03   & 3.5 &  $-$0.08 & 11.13  &   0.06   & \multicolumn{2}{c}{8.21} & 8.33 & 0.11  	    & 8.46 & 0.04	       & \multicolumn{2}{c}{8.43} & \multicolumn{2}{c}{    } & \multicolumn{2}{c}{    } & 7.54 & 0.03		   & 7.46 & 0.04 \\
28 & HD149076   & 11100  & 200  & 2.05  &   0.10  & 5  &   1  & 7   & 37  & 0.56  & 0.02   & 3.5 &  $-$0.44 & 11.14  &   0.05   & \multicolumn{2}{c}{    } & \multicolumn{2}{c}{    } & 8.43 & 0.09	       & 8.44 & 0.10		  & 8.78 & 0.04 	     & \multicolumn{2}{c}{8.83} & \multicolumn{2}{c}{	 } & \multicolumn{2}{c}{    } \\
29 & HD149077   & 9900   & 150  & 2.20  &   0.10  & 3  &   1  & 1   & 13  & 0.53  & 0.03   & 3.5 &  $-$0.20 & 11.14  &   0.06   & \multicolumn{2}{c}{    } & \multicolumn{2}{c}{    } & 8.48 & 0.05	       & \multicolumn{2}{c}{8.41} & \multicolumn{2}{c}{    } & \multicolumn{2}{c}{    } & 7.58 & 0.04		   & 7.51 & 0.04 \\
30 & HD165784   & 9000   & 200  & 1.50  &   0.10  & 7  &   1  & 18  & 35  & 0.86  & 0.02   & 3.1 &  $-$0.02 & 11.13  &   0.03   & \multicolumn{2}{c}{8.41} & 8.37 & 0.05  	    & \multicolumn{2}{c}{    } & \multicolumn{2}{c}{	} & \multicolumn{2}{c}{    } & \multicolumn{2}{c}{    } & 7.58 & 0.08		   & 7.54 & 0.08 \\
31 & HD166167   & 9600   & 150  & 2.00  &   0.10  & 4  &   1  & 9   & 20  & 0.61  & 0.02   & 3.5 &  $-$0.14 & 11.09  &   0.06   & \multicolumn{2}{c}{    } & \multicolumn{2}{c}{    } & \multicolumn{2}{c}{    } & \multicolumn{2}{c}{	} & \multicolumn{2}{c}{    } & \multicolumn{2}{c}{    } & 7.66 & 0.07		   & 7.70 & 0.04 \\
32 & HD80057    & 9300   & 150  & 1.75  &   0.10  & 5  &   1  & 13  & 27  & 0.32  & 0.02   & 3.3 &  $-$0.09 & 11.16  &   0.04   & 8.22 & 0.08		 & 8.28 & 0.13  	    & 8.34 & 0.04	       & \multicolumn{2}{c}{8.29} & \multicolumn{2}{c}{    } & \multicolumn{2}{c}{    } & 7.42 & 0.06		   & 7.40 & 0.06 \\
33 & HD102878   & 8900   & 150  & 1.50  &   0.10  & 6  &   1  & 0   & 35  & 0.27  & 0.02   & 3.4 &  $-$0.02 & 11.15  &   0.05   & 8.26 & 0.12		 & 8.27 & 0.13  	    & \multicolumn{2}{c}{    } & \multicolumn{2}{c}{	} & \multicolumn{2}{c}{    } & \multicolumn{2}{c}{    } & 7.50 & 0.06		   & 7.45 & 0.06 \\
34 & HD105071   & 12000  & 150  & 1.85  &   0.10  & 7  &   1  & 23  & 39  & 0.28  & 0.02   & 3.7 &  $-$0.61 & 11.13  &   0.06   & \multicolumn{2}{c}{    } & \multicolumn{2}{c}{    } & 8.55 & 0.06	       & 8.54 & 0.07		  & 8.78 & 0.04 	     & 8.73 & 0.09		& \multicolumn{2}{c}{	 } & \multicolumn{2}{c}{    } \\
35 & HD106068   & 11600  & 200  & 1.90  &   0.10  & 6  &   1  & 20  & 45  & 0.38  & 0.02   & 3.4 &  $-$0.56 & 11.13  &   0.01   & \multicolumn{2}{c}{    } & \multicolumn{2}{c}{    } & 8.60 & 0.04	       & 8.60 & 0.04		  & 8.75 & 0.04 	     & 8.82 & 0.04		& \multicolumn{2}{c}{	 } & \multicolumn{2}{c}{    } \\[1mm]
\hline
\end{tabular}
\tablefoot{
\tablefoottext{a}{Typical uncertainties are $\pm$5\,km\,s$^{-1}$, and 
$\pm$3\,km\,s$^{-1}$ for the stars with the sharpest lines.}
\tablefoottext{b}{Typical uncertainties are $\pm$0.2.}
\tablefoottext{c}{Typical uncertainties vary from $\sim$0.02\,mag for the 
coolest sample stars to $\sim$0.05\,mag for the hottest stars.}
\tablefoottext{d}{In cases where only one spectral line could be analysed 
for an ion (no error margin given), a good estimate for
the uncertainty is 0.10\,dex.}
}
\end{table}
\end{landscape}
\noindent

\begin{table}[t!]
\centering
 \setlength{\tabcolsep}{.1cm}
\caption{Comparison of our atmospheric parameters
($T_{\mathrm{eff}}$\,[K], $\log g$) for objects in common 
with previous studies.\label{compare}}
 \begin{tabular}{rlcccr}
 \noalign{}
\hline\hline
\#  & Object  & Takeda\tablefootmark{a}  & Venn\tablefootmark{b} &
McErlean\tablefootmark{c} & this work\\
\hline\\[-3mm]
1  & HD12301    & 12500, 2.30 & \ldots      &  14000, 2.15 &   12600, 2.15\\ 
3  & HD13476    &  9000, 1.50 &  8400, 1.20 &  \ldots	   &    8500, 1.40\\ 
6  & HD14489    & \ldots      &  9000, 1.40 &  \ldots	   &    9350, 1.45\\ 
8  & HD21291    & 12000, 1.80 &  \ldots     &  11500, 1.60 &   10800, 1.65\\ 
9  & HD39970    & 10500, 1.70 &  \ldots     &  \ldots	   &   10300, 1.70\\ 
10 & HD46300    & 10000, 2.00 &  9700, 2.10 &  \ldots	   &   10000, 2.15\\ 
13 & HD197345   &  9500, 1.50 &  \ldots     &  \ldots	   &    8700, 1.20\\ 
14 & HD202850   & 11500, 1.80 &  \ldots     &  \ldots	   &   10800, 1.85\\ 
15 & HD207260   &  9500, 1.50 &  \ldots     &  \ldots	   &    8800, 1.35\\ 
16 & HD207673   & \ldots      &  9300, 1.75 &  \ldots	   &    9250, 1.80\\
17 & HD208501   & 12500, 2.30 &  \ldots     &  13000, 1.80 &   12700, 1.85\\ 
18 & HD210221   & \ldots      &  8200, 1.30 &  \ldots	   &    8400, 1.40\\ 
19 & HD212593   & \ldots      &  \ldots     &  \ldots	   &   11200, 2.10\\ 
23 & HD195324   & \ldots      &  9300, 1.90 &  \ldots	   &    9200, 1.85\\
24 & HD34085    & 13000, 2.00 &  \ldots     &  13000, 1.75 &   12100, 1.75\\ 
25 & HD87737    & 10200, 1.90 &  9700, 2.00 &  \ldots	   &    9600, 2.05\\ 
\hline\\[-5mm]
\end{tabular}
\tablefoot{
\tablefoottext{a}{\citet{Takeda};}
\tablefoottext{b}{\citet{Venn95a};}
\tablefoottext{c}{\citet{McErlean}.}
}
\end{table}

Our accuracy in microturbulence determination is limited by the stepsize 
of 1\, km\,s$^{-1}$ in our model grid. The values derived from different 
elemental species are consistent within this error. The derived helium 
abundances vary between 10 and 14\% in number fraction, with errors of 
around 20\% in these values. Note, however, that the high sensitivity of the 
helium lines to $T_{\mathrm{eff}}$-variations makes the values prone to 
substantial systematic shifts even from small errors in the parameter 
determination. Finally, metal abundances are determined on average
with about 10--20\% uncertainty. Abundances from lines of the major ionization
stages are usually insensitive to variations of $T_{\mathrm{eff}}$ and
$\log g$ within the given uncertainties.

\subsection{Comparison with previous analyses}
Samples of about 10--20 Galactic supergiants in the BA-star 
regime were subject to non-LTE
analyses in three previous studies \citep{Venn95a,McErlean,Takeda}. 
We list their effective temperatures and surface gravities for stars in common 
with the present work in Table~\ref{compare}. Typical uncertainties
in $T_\mathrm{eff}$/$\log g$ are 200\,K/0.2\,dex,
500--1000\,K/0.5\,dex and 1000\,K/0.2\,dex in these studies, respectively.

\citet{Venn95a} used a similar method for the 
parameter determination in A-type supergiants, i.e., utilising the Balmer- and 
Mg\,{\sc i/ii}-lines, however without a simultaneous derivation of the
He abundance in her hybrid non-LTE approach. While there are some 
differences, good overall agreement is 
found in $T_\mathrm{eff}$ and $\log g$ within the uncertainties. 

\citet{Takeda} tried to construct a $T_{\mathrm{eff}}/\log g$ vs. spectral type 
relation, utilising published data. They emphasised, however, that this approach 
is subject to considerable uncertainties. Indeed, we find notable
deviations between our values for both $T_{\mathrm{eff}}$ and $\log g$
and their data, though agreement is still found when considering their
rather large uncertainty ranges.

Finally, there is some overlap with the B-type supergiants 
studied by \citet{McErlean}. Their estimates for $\log g$, inferred from the 
Balmer lines, are in excellent agreement with ours. Yet, we find a systematic 
shift in $T_{\mathrm{eff}}$ towards higher values compared to our results. 
Their temperatures are based on the non-LTE Si\,{\sc ii/iii} ionization balance for 
the stars in common (except for HD21291, where photometry was used 
as substitute). However, the non-LTE model atmospheres used for their
analysis did not consider the effects of metal line-blanketing, which
we identify as the likely cause of the $T_{\mathrm{eff}}$-differences.

\section{First applications}

\begin{table*}[t]
\centering
\setlength{\tabcolsep}{.14cm}
\caption{Spectral type--$T_\mathrm{eff}$--$(B-V)_0$--$(b-y)_0$ scales \& $B.C.$ values 
for Galactic BA-type supergiants.\label{scales}}
 \begin{tabular}{crrrrrrcrrrr}
 \noalign{}
\hline\hline
        & \multicolumn{6}{c}{reference values from literature} & &
	\multicolumn{4}{c}{this work\tablefootmark{e}}\\
\cline{2-7} \cline{9-12}
Sp.\,T. & $T_\mathrm{eff}$\tablefootmark{a} & $(B-V)_0$\tablefootmark{a} & 
$T_\mathrm{eff}$\tablefootmark{b} & $(B-V)_0$\tablefootmark{b} 
& $(b-y)_0$\tablefootmark{c}
& $B.C.$\tablefootmark{a,b,d} & & $T_\mathrm{eff}$ & $(B-V)_0$ & 
$(b-y)_0$ & $B.C.$\\
        & K & mag & K & mag& mag & mag & & K & mag & mag & mag\\
\hline\\[-3mm]
B6 & 13000 & $-$0.08 & {\ldots} & {\ldots} &        {\ldots} & $-$0.76  & & {\ldots}       & {\ldots}         & {\ldots} & {\ldots}\\
B7 & 12200 & $-$0.05 & {\ldots} & {\ldots} &        {\ldots} & $-$0.66	& & {\ldots}       & {\ldots}         & {\ldots} & {\ldots}\\
B8 & 11200 & $-$0.03 & 11100    &  $-0.03$ &        {\ldots} & $-$0.54	& & 12200$\pm$410  & $-$0.09$\pm$0.01 & 0.001$\pm$0.009 & $-$0.66$\pm$0.08\\
B9 & 10300 & $-$0.02 & {\ldots} & {\ldots} &        {\ldots} & $-$0.40	& & 10920$\pm$220  & $-$0.06$\pm$0.02 & 0.011$\pm$0.012 & $-$0.31$\pm$0.04\\
A0 &  9730 & $-$0.01 &  9980    &  $-0.01$ & 0.032$\pm$0.019 & $-$0.29	& &  9840$\pm$290  & $-$0.05$\pm$0.01 & 0.017$\pm$0.013 & $-$0.19$\pm$0.05\\
A1 &  9230 &    0.02 & {\ldots} & {\ldots} &        {\ldots} & $-$0.20	& &  9240$\pm$~~80 & $-$0.01$\pm$0.02 & 0.043$\pm$0.014 & $-$0.09$\pm$0.03\\
A2 &  9080 &    0.03 &  9380    &     0.03 & 0.051$\pm$0.011 & $-$0.16	& &  8960$\pm$200  &    0.00$\pm$0.02 & 0.051$\pm$0.015 & $-$0.03$\pm$0.05\\
A3 &  8770 &    0.06 & {\ldots} & {\ldots} & 0.074$\pm$0.021 & $-$0.09	& &  8430$\pm$~~60 &    0.01$\pm$0.01 & 0.056$\pm$0.005 & $+$0.08$\pm$0.02\\
A5 &  8510 &    0.09 &  8610    &     0.09 & 0.062$\pm$0.017 & $-$0.01  & & {\ldots}       & {\ldots}         & {\ldots} & {\ldots}\\
\hline\\[-5mm]
 \end{tabular}
\tablefoot{
\tablefoottext{a}{\citet{SK};}
\tablefoottext{b}{\citet{Cox};}
\tablefoottext{c}{\citet{Gray92}, for LC Ib;} 
\tablefoottext{d}{corrected by $+$$0\,{\fm}12$ to match the
zeropoint of the present work;}
\tablefoottext{e}{see Sects.~\ref{sptteffsect}--\ref{bolcorrections} 
for a discussion.}
}
\end{table*}

\begin{figure}
\centering
\includegraphics[width=.99\linewidth]{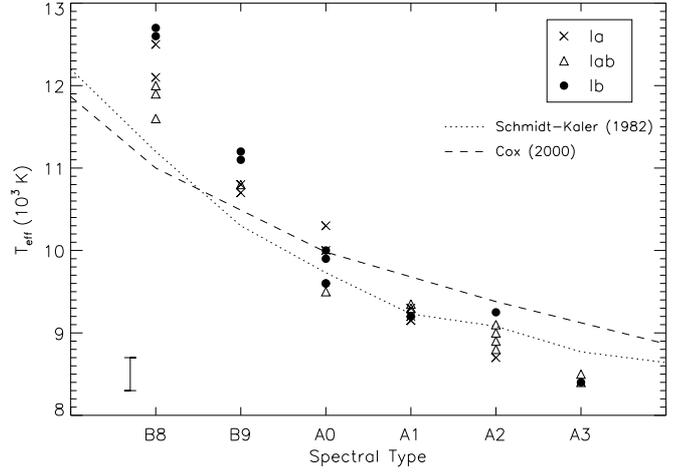}
\caption{Comparison of our results for the individual sample stars
with reference spectral type--$T_\mathrm{eff}$ scales. 
Luminosity classes and reference scales are encoded according to the legend. 
A typical error bar is indicated to the lower left. See
Sect.~\ref{sptteffsect} for a discussion.}
\label{sptteff}
\end{figure}

\subsection{Spectral type--T$_{\sf eff}$ relation \label{sptteffsect}}

Empirical spectral-type--$T_\mathrm{eff}$ relations provide important
starting points for all kinds of stellar studies and for
quantitative spectroscopy in particular, and they are therefore 
an essential part of the {\em reference literature} on stellar
properties. Our high-precision/high-accuracy dataset facilitates to 
reassess the existing knowledge in the BA-type supergiant regime in view 
of improved models and analysis techniques. 

Effective temperatures of our sample stars (see Table~\ref{parameters}) as a 
function of their
spectral type (according to the refined classification presented in 
Table~\ref{observations1}) are displayed in Fig.~\ref{sptteff}.
In comparison with established reference work \citep{SK,Cox}
we find a significantly steeper spectral-type--$T_\mathrm{eff}$ relation, i.e. 
higher $T_\mathrm{eff}$ at spectral types B8 and B9, and lower 
$T_\mathrm{eff}$ at A2 and A3. As no apparent correlation of 
$T_\mathrm{eff}$ with luminosity class (LC) is indicated in
Fig.~\ref{sptteff}, we compute average $T_\mathrm{eff}$-values
from all stars of a spectral subtype to provide refined reference
values which are presented in Table~\ref{scales}. 

With typically five
to seven objects per spectral type it would be desirable to increase 
the sample size in order to get a more robust determination, in
particular of the uncertainties. The distribution of the data per 
spectral bin is not Gaussian at the moment. Note that the formal 
1$\sigma$-uncertainties for spectral types A1 and A3 in
Table~\ref{scales} are lower than the uncertainties from the
individual star analyses. There is little
overlap of $T_\mathrm{eff}$-values for individual stars of different
spectral types. Overall it appears that our refined 
spectral-type--$T_\mathrm{eff}$ relation allows effective temperatures to 
be estimated with a typical 1$\sigma$-uncertainty of about 2--3\%.

\begin{figure}
\centering
\includegraphics[width=.99\linewidth]{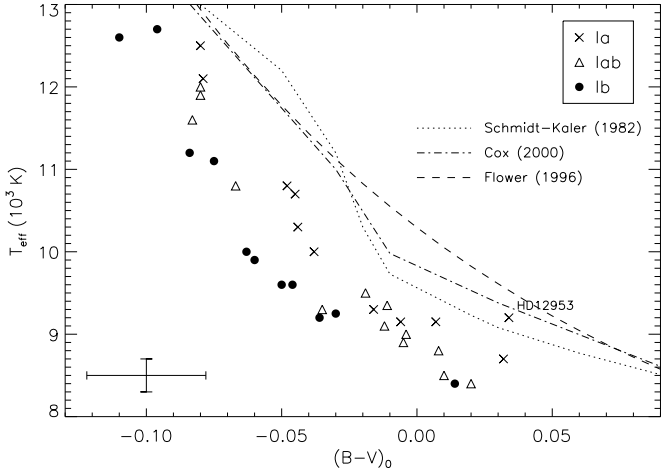}
\caption{Comparison of observational results for the individual sample stars
with established reference $(B-V)_0$--$T_\mathrm{eff}$ relations from the
literature, according to the legend.
A typical error bar is indicated.
See Sect.~\ref{bmvscale} for a discussion.}
\label{bminusv}
\end{figure}

Obviously, the work has to be extended in the future to provide 
a unified picture accounting for
recent developments concerning hotter \citep[e.g.][]{Repolust04,MaPu08} and 
cooler Galactic supergiants \citep[e.g.][]{Levesque05}. 

\subsection{(B--V)$_{\sf 0}$--T$_{\sf eff}$ scale}\label{bmvscale}
Closely related to the spectral-type--$T_\mathrm{eff}$ relation is the
question of the behaviour of intrinsic Johnson colour $(B-V)_0$ with
spectral type, or more precisely, with effective temperature.
We determined $(B-V)_0$ from the observed $(B-V)$ colour (see
Table~\ref{observations1}), corrected for the colour excess 
$E(B-V)$ according to Table~\ref{parameters}. 

Our effective temperatures
as a function of $(B-V)_0$ are displayed in Fig.~\ref{bminusv}, together 
with established reference relations from \citet[for LC\,Iab]{SK}, 
\citet[LC\,I]{Flower96} and \citet[LC\,I]{Cox}. Our data are
systematically bluer by $0\fm03$--$0\fm06$ than indicated by the
reference relationships. The supergiants become bluer with decreasing
luminosity (i.e.~smaller radius) for a given $T_\mathrm{eff}$, the
decrement being about $0\fm01$ per luminosity subclass. As we do not
have enough objects per spectral type to compute meaningful means
for the individual luminosity subclasses, we provide $(B-V)_0$-values 
averaged over the entire LC\,I in Table~\ref{scales}. Note that the outlier
HD\,12953 -- the by far the most luminous 
star of our sample -- has been excluded from this.

\begin{figure}
\centering
\includegraphics[width=.99\linewidth]{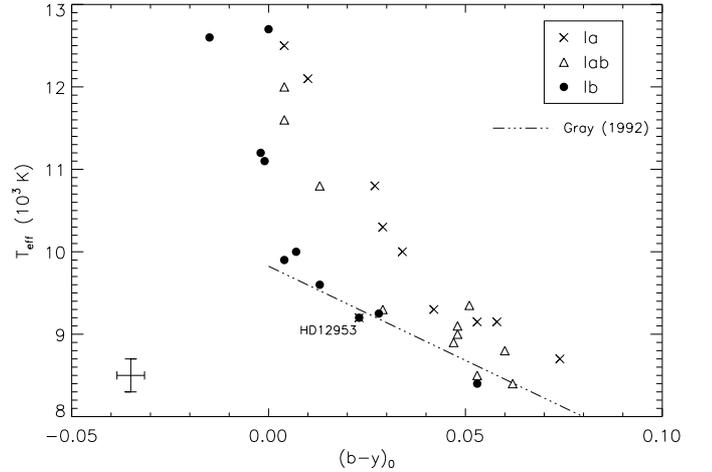}
\caption{Comparison of observational results for the sample stars
with a reference $(b-y)_0$--$T_\mathrm{eff}$ relation from the
literature, according to the legend. A typical error bar is indicated.
See Sect.~\ref{bmyscale} for a discussion.}
\label{bminusy}
\end{figure}

The loci of the three luminosity subclasses of BA-type supergiants in
Fig.~\ref{bminusv} may be approximated by a quadratic fit function, 
yielding the following relations\\[2mm]
$~~~~~~~~~(B-V)_0\,(\mathrm{mag})=$\\[-.5cm]
\begin{eqnarray}
\label{bmvIacal}\mathrm{Ia:}   & ~50.668-24.585\,\log
T_\mathrm{eff}+2.977\,(\log T_\mathrm{eff})^2\\
\label{bmvIabcal}\mathrm{Iab:} & ~34.368-16.529\,\log
T_\mathrm{eff}+1.981\,(\log T_\mathrm{eff})^2\\
\label{bmvIbcal}\mathrm{Ib:}   & ~49.565-24.114\,\log
T_\mathrm{eff}+2.927\,(\log T_\mathrm{eff})^2
\end{eqnarray}
with an area of validity of about 
$3.92\lesssim \log T_\mathrm{eff} \lesssim 4.10$.
Like for the case of the spectral type--$T_\mathrm{eff}$ relation it
would be desirable to extend this to hotter and cooler supergiants. 

\subsection{(b--y)$_{\sf 0}$--T$_{\sf eff}$ scale}\label{bmyscale} 
Our effective temperatures as a function of Str\"omgren $(b-y)_0$ 
are displayed in Fig.~\ref{bminusy}, together 
with a reference relation from \citet[for LC\,Ib]{Gray92}.
There is an excellent match of Gray's relation with our data. Note
the drastic change of the slope of the
$(b-y)_0$--$T_\mathrm{eff}$-relation for bluer $(b-y)_0$. In general, 
the supergiants become bluer with decreasing
luminosity for a given $T_\mathrm{eff}$, the
decrement being about $0\fm01$ per luminosity subclass. Like for the
previous case of the Johnson data, we provide $(b-y)_0$-values
averaged over the entire LC\,I in Table~\ref{scales}. The outlier      
HD\,12953, which appears exceptionally {\em blue} in $(b-y)_0$, has
been excluded.

\begin{figure}
\centering
\includegraphics[width=.99\linewidth]{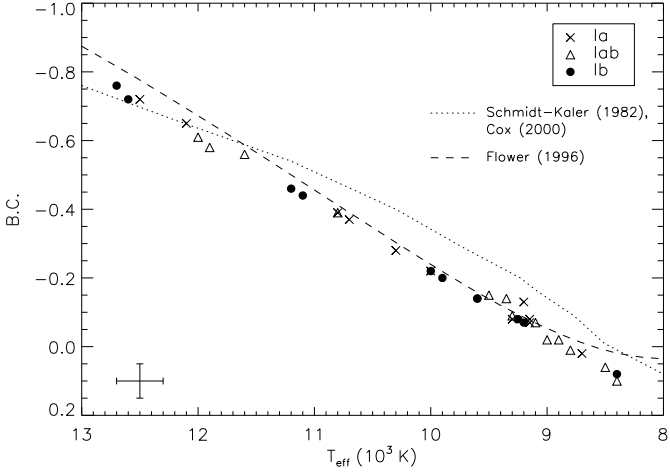}
\caption{Comparison of bolometric corrections determined here 
with reference relations from the
literature, according to the legend.
The error bar denotes a typical uncertainty in $T_\mathrm{eff}$ and a
{\em maximum} uncertainty in $B.C.$.
See Sect.~\ref{bolcorrections} for a discussion.}
\label{bolcorr}
\end{figure}

The loci of the three luminosity subclasses of BA-type supergiants in
Fig.~\ref{bminusy} may be approximated by a quadratic fit function, 
yielding the following relations\\[2mm]
$~~~~~~~~~(b-y)_0\,(\mathrm{mag})=$\\[-.5cm]
\begin{eqnarray}
\label{bmyIacal}\mathrm{Ia:}   & ~30.222-14.633\,\log
T_\mathrm{eff}+1.771\,(\log T_\mathrm{eff})^2\\
\label{bmyIabcal}\mathrm{Iab:} & ~12.147-~\,5.669\,\log
T_\mathrm{eff}+0.659\,(\log T_\mathrm{eff})^2\\
\label{bmyIbcal}\mathrm{Ib:}   & ~42.265-20.736\,\log
T_\mathrm{eff}+2.543\,(\log T_\mathrm{eff})^2
\end{eqnarray}
with an area of validity of about 
$3.92\lesssim \log T_\mathrm{eff} \lesssim 4.10$.

\subsection{Bolometric corrections}\label{bolcorrections}
We determined bolometric corrections from the model fluxes for each 
individual star. Following the approach of \citet{Bessell}, 
the model flux at the $V$-band {\em effective}
wavelength was converted into a Johnson $V$ magnitude in the usual
way by adopting a zeropoint flux 
$f_{\nu}^0$\,=\,$3.636 \times 10^{-20}\,\mbox{erg
cm}^{-2}\,\mbox{s}^{-1}\,\mbox{Hz}^{-1}$ (for Vega).
The integral over the total model spectral energy distribution
yielded a bolometric magnitude (assuming an absolute solar             
bolometric magnitude $M_\mathrm{bol}^{\odot}$\,=\,$+4.74$).
The difference between the bolometric and the $V$ magnitudes calculated 
in this way then provided the bolometric correction, which can be
expressed as 
\begin{equation}
B.C.= M_\mathrm{bol}-M_V = C - 10
\log (T_{\mathrm{eff}}/T_\mathrm{eff}^{\odot})+2.5 
\log{H_{\nu}}\,,
\label{bceqn}
\end{equation}
where $C$\,=\,12.854 for a solar radius $R_{\odot}$\,=\,$6.95 \times 
10^{10}\,\mbox{cm}$ and $T_\mathrm{eff}^{\odot}$\,=\,5777\,K, and 
$H_{\nu}$ is the Eddington flux at the effective $V$-band frequency. 
This way, the correct empirical $B.C.$ for the standard star Vega is 
reproduced, which lies exactly in the $T_\mathrm{eff}$-range 
spanned by the sample stars. Note that for this reason we have not 
renormalised the bolometric corrections to the minimum value of zero
in order to avoid positive $B.C.$-values for some late-A and F-type stars, as
suggested by \citet{BuKu78}, see also the discussion of this topic by
\citet{Bessell} and \citet{Torres10}. As a result, our $B.C.$ values are
about $0\,{\fm}12$ larger than the ones provided earlier by us 
when introducing the modelling techniques and analysis methodology
used here \citep{Norbert}.
The $B.C.$-values determined from Eq.~\ref{bceqn}, summarised in
Table~\ref{parameters}, deviate less than 
$0\,{\fm}03$ from the recent analytical fit formula of
\citet[their Eq.~6]{Kudritzki08}, which have used the same approach.
The uncertainties of the bolometric corrections are dominated by
the $T_\mathrm{eff}$-errors. For the coolest sample stars, i.e.~close 
to the minimum of the $T_\mathrm{eff}$--$B.C.$-curve, the uncertainties 
for individual objects amount to $\sim$0.02\,mag, while they reach up 
to $\sim$0.05\,mag for maximum $T_\mathrm{eff}$-uncertainties in 
the hottest stars.

The derived values are compared to the reference calibrations of
\citet[for stars of luminosity class Iab]{SK}, which are reproduced by
\citet{Cox}, and of \citet[for
supergiants, his Table~4]{Flower96}
in Fig.~\ref{bolcorr}. In order to
adjust for the different zeropoints used, we correct the data of
Schmidt-Kaler/Cox by 
$0\,{\fm}12$\footnote{\citet{SK} adopts 
$M_\mathrm{bol}^{\odot}$\,=\,$-0\,{\fm}19$, while the value provided 
by \citet{Bessell} is $-0\,{\fm}07$. Note that the Schmidt-Kaler data is 
subject to a further internal inconsistency by $0\,{\fm}02$ 
\citep{Torres10}, which would
improve the comparison with our values only slightly.}. The 
resulting differences can reach values up to $0\,{\fm}13$. A general trend 
towards larger $B.C.$-values is seen in our data, except for the hottest 
objects in the sample, which have slightly lower bolometric corrections.
The match of the Flower relation (adjusted by $+0\fm01$ because of the
different zeropoints) on the other hand is good, with small deviations
becoming apparent at the highest and lowest temperatures considered here.

As no significant deviations occur due to luminosity classification,
except possibly for our most luminous sample star HD\,12953, we average
over all members of a spectral subtype to determine improved $B.C.$
reference values. Our values are compared to the reference data from
the literature in Table~\ref{scales}. 

\subsection{Photometric atmospheric parameter determination}
As one often wishes to analyse larger samples of stars, easy-to-apply
and fast stellar parameter indicators are in high demand. Photometric 
$T_\mathrm{eff}$-indicators are among the preferred ones, because of
the easy accessibility of photometric data. We discuss empirical
relations for the two most common photometric systems, 
for Johnson and for Str\"omgren photometry. The latter also allows
surface gravities to be estimated photometrically.

\paragraph{The reddening-free Johnson Q-index as T$_{\sf eff}$-indicator.}
Effective temperature calibrations based on the reddening-free $Q$-index
\citep{Johnson58}, $Q$\,=\,$(U-B)-X(B-V)$ with $X$\,=\,$E(U-B)/E(B-V)$, 
have come into wide use recently, as Johnson photometry is available
for most stars. Examples in our context encompass studies of the
evolutionary progenitors of BA-type supergiants, OB-type stars on the main sequence 
\citep[e.g.][]{DaCu99,Lyubimkov02}, and their cooler siblings, late-A to
G-type supergiants \citep{Lyubimkov}. Usually, a standard value of 0.72
is adopted for $X$, but see \citet{Johnson58}. 

\begin{figure}
\centering
\includegraphics[width=.99\linewidth]{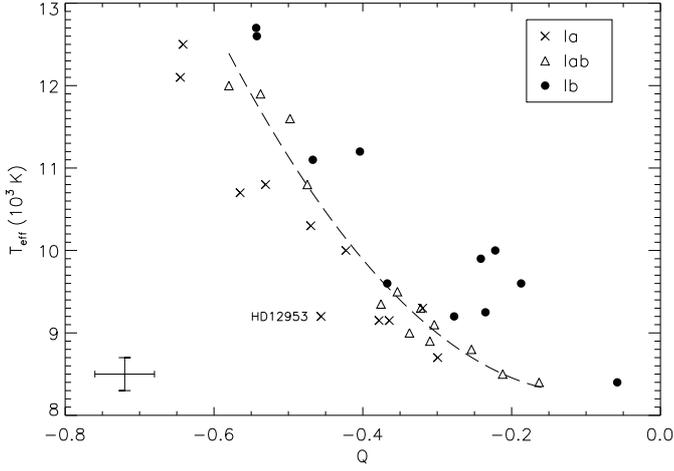}
\caption{Relation between the reddening-free Johnson $Q$-index,
$Q$\,$=$\,$(U-B)-0.72(B-V)$, and the spectroscopic 
$T_{\mathrm{eff}}$-values of the sample stars. A typical error bar is 
indicated. The dashed line represents the regression line of our
$Q$-based $T_\mathrm{eff}$-calibration for Iab supergiants. 
}
\label{qfactor}
\end{figure}

We test the applicability of the $Q$-index as possible
$T_{\mathrm{eff}}$-indicator for BA-type supergiants in
Fig.~\ref{qfactor}. Our spectroscopically derived
$T_\mathrm{eff}$-values are displayed as a function of $Q$. A trend
with luminosity subclass becomes apparent, the more luminous Ia objects
being coolest and the Ib objects being hottest at a given $Q$. The
loci of the three luminosity subclasses in the diagram may be
approximated by a quadratic fit function, yielding the following
relations\\[2mm]
$~~~~~~~~~T_\mathrm{eff}^Q\,(10^3\,\mathrm{K})=$\\[-.5cm]
\begin{eqnarray}
\label{Iacal}\mathrm{Ia:}   & ~8.691+3.447\,Q+13.843\,Q^2\,,\,~-0.30\gtrsim Q\gtrsim-0.65\\
\label{Iabcal}\mathrm{Iab:} & ~8.435+3.430\,Q+17.672\,Q^2\,,~~-0.15\gtrsim Q\gtrsim -0.60\\
\label{Ibcal}\mathrm{Ib:}   & ~8.725-0.026\,Q+12.631\,Q^2\,,~~-0.05\gtrsim Q\gtrsim -0.55
\end{eqnarray}
with their respective area of validity. The outlier HD\,12953 was
excluded from the fitting procedure. For clarity, only the fit
function for Iab supergiants is visualised in Fig.~\ref{qfactor}. 
At first glance, rather tight relations are found.

\begin{figure}
\centering
\includegraphics[width=.99\linewidth]{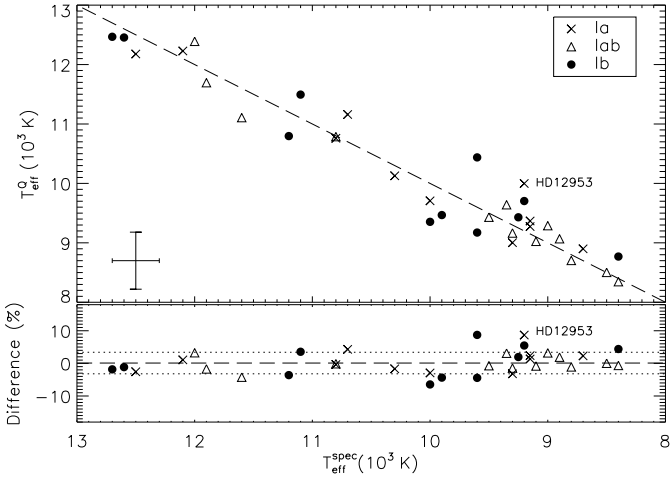}
\caption{Upper panel: comparison of our spectroscopically derived
$T_\mathrm{eff}^\mathrm{spec}$ with the $Q$-based $T_\mathrm{eff}^Q$
computed with Eqs.~\ref{Iacal}--\ref{Ibcal}. The dashed line
indicates the 1:1 relation. The error bar shows a typical conservative 
uncertainty from our spectroscopic determination, and the uncertainty
in $T_\mathrm{eff}^Q$, accounting for {\em typical uncertainties in the
colours only}. Lower panel: percent difference of the two
$T_\mathrm{eff}$-values for the individual objects. These quantify
the {\em systematic differences} from application of our
$Q$--$T_\mathrm{eff}$ calibrations. The dotted lines indicate the
1$\sigma$-scatter range.}
\label{teffteff}
\end{figure}

The differences between the spectroscopically derived effective
temperatures $T_\mathrm{eff}^\mathrm{spec}$ and the $Q$-based
$T_\mathrm{eff}^Q$ computed with Eqs.~\ref{Iacal}--\ref{Ibcal}
are quantified in Fig.~\ref{teffteff}. The sample stars follow the 1:1
relation rather tightly, with the 1$\sigma$-scatter of the computed
$T_\mathrm{eff}^Q$ around the spectroscopic reference values amounting
to more than 3\%. This is the {\em systematic uncertainty}
adherent to an application of the method. In addition, {\em random
errors} because of the uncertainties in the colours also have to be
considered, typically amounting to more than twice the 1$\sigma$-uncertainty
of the spectroscopically derived values, i.e.~another $\sim$5\% 
(see the error bar in the
upper panel of Fig.~\ref{teffteff} and note the different scale
projection of the axes). In consequence, {\em the resulting error margins
render this simple approach of $Q$-based effective temperatures not 
competitive with spectroscopic determinations for BA-type supergiants}. 
Moreover, better starting points for an iterative refinement of the 
stellar parameters are already obtained from the spectral 
type--$T_\mathrm{eff}$ relation established in Sect.~\ref{sptteffsect}.

\begin{figure}
\centering
\includegraphics[width=.99\linewidth]{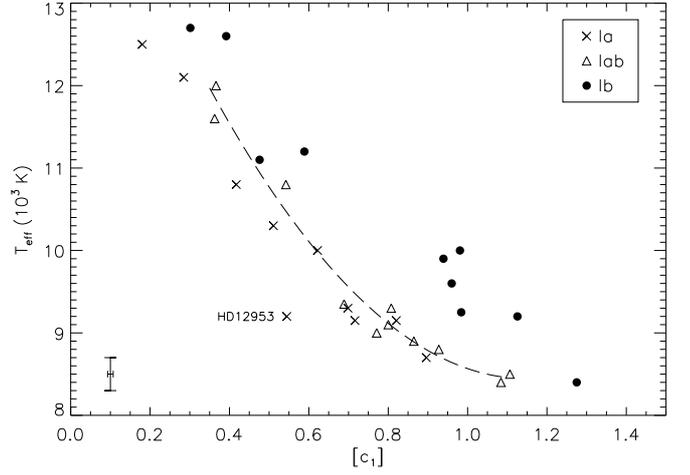}
\caption{As Fig.~\ref{qfactor}, but for the
relation between the reddening-free Str\"omgren $[c_1]$-index,
$[c_1]$\,$=$\,$c_1-0.20(b-y)$  where $c_1=(u-v)-(v-b)$),
and the spectroscopic                                
$T_{\mathrm{eff}}$-values of the sample stars.}
\label{c1factor}
\end{figure}

\begin{figure}
\centering
\includegraphics[width=.99\linewidth]{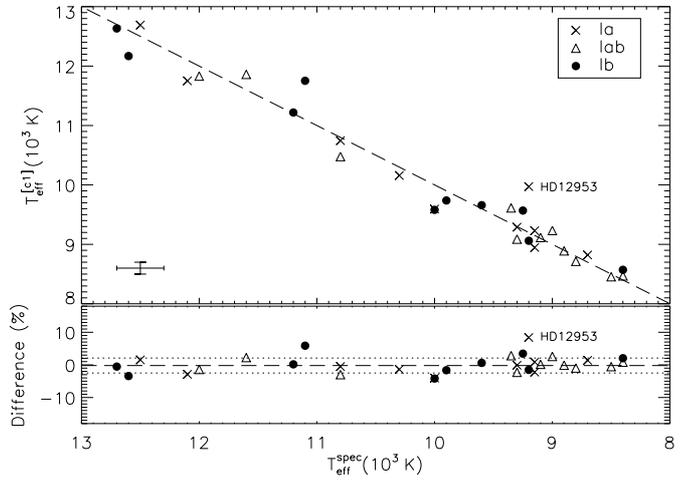}
\caption{As Fig.~\ref{teffteff}, but for the comparison of our 
spectroscopically derived
$T_\mathrm{eff}^\mathrm{spec}$ with the $[c_1]$-based
$T_\mathrm{eff}^{[c_1]}$
computed with Eqs.~\ref{Iacalstr}--\ref{Ibcalstr}.}
\label{teffteffstr}
\end{figure}

\paragraph{The reddening-free Str\"omgren [c$_{\sf 1}$]-index as 
T$_{\sf eff}$-indicator.}
Data for 31 of our sample supergiants were found in the catalogue of 
\citet{HauMer98} on Str\"omgren $uvby\beta$ photometry,
facilitating the reddening-free $[c_1]$-index 
$[c_1]$\,$=$\,$c_1-0.20(b-y)$ to be computed, with
$c_1=(u-v)-(v-b)$. Figure~\ref{c1factor} 
visualises the sample stars in the $[c_1]$-$T_\mathrm{eff}$ plane. The
loci of the three luminosity subclasses may be approximated by a quadratic 
fit function as before, yielding the following relations\\[2mm]
$~~~~~~~~~T_\mathrm{eff}^{[c_1]}\,(10^3\,\mathrm{K})=$\\[-.5cm]
\begin{eqnarray}
\label{Iacalstr}\mathrm{Ia:}   & 14.610-11.706\,[c_1]+5.855\,[c_1]^2\,,0.15\lesssim [c_1]\lesssim0.90\\
\label{Iabcalstr}\mathrm{Iab:} & 15.741-12.710\,[c_1]+5.536\,[c_1]^2\,,0.35\lesssim [c_1]\lesssim 1.10\\
\label{Ibcalstr}\mathrm{Ib:}   & 14.310-~\,5.893\,[c_1]+1.092\,[c_1]^2\,,0.25\lesssim [c_1]\lesssim 1.25
\end{eqnarray}
with their respective area of validity. For clarity, only the fit
function for Iab supergiants is visualised in Fig.~\ref{c1factor}.
The clear outlier HD\,12953 was ignored in the
determination of the fit coefficients in Eq.~\ref{Iacalstr}. The
Str\"omgren $[c_1]$-$T_\mathrm{eff}$ relations are obviously tighter than the
Johnson $Q$-$T_\mathrm{eff}$ relations. In fact, a practical common relation for
BA-type supergiants of luminosity class Ia {\em and} Iab can be given
with only a negligible loss in accuracy for the range 
$0.15 \lesssim [c_1] \lesssim 1.10$:
\begin{eqnarray}
\label{IaIabcalstr}
T_\mathrm{eff}^{[c_1]}\,(10^3\,\mathrm{K})=14.613-10.565\,[c_1]+4.523\,[c_1]^2\,.
\end{eqnarray}

\begin{figure}
\centering
\includegraphics[width=.99\linewidth]{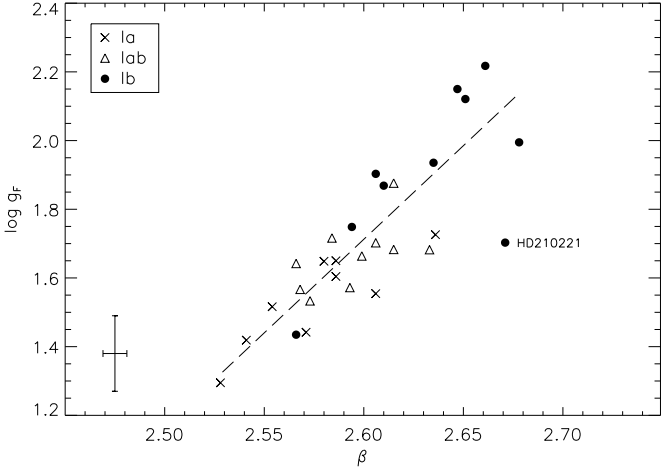}
\caption{Relation between Str\"omgren $\beta$ and the
spectroscopically derived flux-weighted
gravities $\log g_\mathrm{F}$ of the sample stars. The error bar in
$\beta$ is an estimate based on the restricted information available
from the Str\"omgren data. The dashed line represents the linear
regression to the data, ignoring the outlier HD\,210221.}
\label{betaloggf}
\end{figure}

In analogy to Fig.~\ref{teffteff}, the differences between
$T_\mathrm{eff}^\mathrm{spec}$ and the $[c_1]$-based
$T_\mathrm{eff}^{[c_1]}$ computed with Eqs.~\ref{Iacalstr}--\ref{Ibcalstr}
are quantified in Fig.~\ref{teffteffstr}. The 1$\sigma$-scatter of
the computed $T_\mathrm{eff}^{[c_1]}$ around the spectroscopic reference
values amounts to about 2\% only in this case (systematic uncertainty
of the method). The random errors due to Str\"omgren colour uncertainties 
are also drastically reduced compared to Johnson photometry, amounting to 
about 1\%. 
We conclude that {\em the reddening-free $[c_1]$-index 
provides highly useful starting values for $T_\mathrm{eff}$ in
high-precision/accuracy analyses of all except the most luminous
BA-type supergiants}. 

As the reddening-free Str\"omgren $[c_1]$-index measures the Balmer
decrement\footnote{Note that the Balmer jump shows some sensitivity to
variations of surface gravity as well.} at the temperatures of the 
sample stars, it could act as a substitute
for the Balmer jump method \citep{Kudritzki08} based on 
flux-calibrated low/intermediate spectroscopy, that has been used 
for studies of BA-type supergiants in galaxies beyond the Local Group
recently. Extended exposures in a spectral region with usually low
instrumental sensitivity (decrease of mirror coating reflectivity and
CCD sensitivity towards the optical-UV) could thus be avoided.
However, practical challenges to be solved first are the
determination of the metallicity dependency of the
$[c_1]$-$T_\mathrm{eff}$ relation and the availability of Str\"omgren
filter sets at large telescopes.

\begin{figure}
\centering
\includegraphics[width=.99\linewidth]{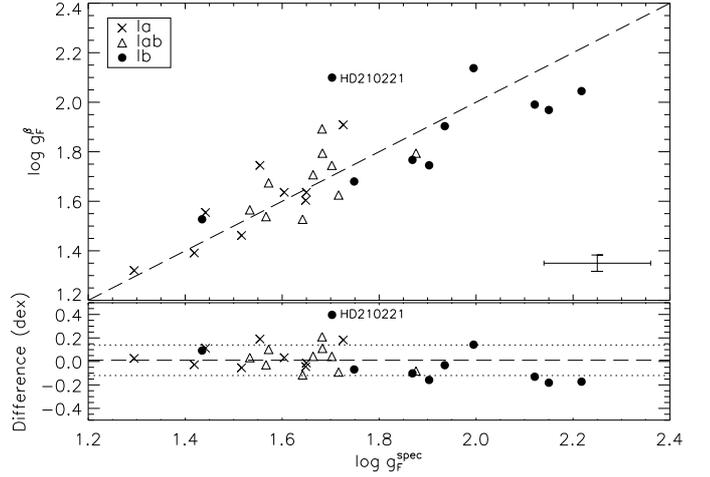}
\caption{Upper panel: comparison of our spectroscopically derived
$\log g_\mathrm{F}^\mathrm{spec}$ with the $\beta$-based 
$\log g_\mathrm{F}^{\beta}$ computed with Eq.~\ref{gfcal}. 
The dashed line indicates the 1:1 relation. The error bar
shows a typical uncertainty from our spectroscopic
determination, and the uncertainty in $\log g_\mathrm{F}^{\beta}$, 
accounting for {\em typical uncertainties in $\beta$ only}. 
Lower panel: difference of the two $\log g_\mathrm{F}$-values for 
the individual objects, in dex. These quantify the 
{\em systematic differences} from application of our                         
$\beta$--$\log g_\mathrm{F}$ calibration. The dotted
lines indicate the 1$\sigma$-scatter range.}
\label{loggfloggf}
\end{figure}

\paragraph{Str\"omgren $\beta$ as a photometric gravity indicator.}
The Str\"omgren $\beta$-index is a poor {\em direct} indicator for 
surface gravity in the parameter range of BA-type supergiants 
because the strength of the Balmer lines depends not
only on $\log g$ -- pressure broadening via the linear Stark effect -- alone
but also on $T_\mathrm{eff}$ -- via (de-)population of the $n$\,=\,2
level of neutral hydrogen by excitation/ionization. On the other hand, 
there is a good correlation with the flux-weighted gravity
\begin{equation}
\log g_\mathrm{F}=\log g - 4 \log T_\mathrm{eff,4}\,,
\end{equation}
as illustrated by Fig.~\ref{betaloggf}. Here,
$T_\mathrm{eff,4}$\,=\,$T_\mathrm{eff}$/10\,000\,K. The flux-weighted
gravity is a measure for the stellar luminosity 
\citep{Kudritzki03,Kudritzki08}, with lower values indicating higher
luminosity. Linear regression of the data yields
\begin{eqnarray}
\label{gfcal}
\log g_\mathrm{F}^{\beta}=-12.4548+5.449\,\beta\,.
\end{eqnarray}
One statistical outlier, HD\,210221, was excluded from the fit 
based on its distance to the remainder of the data points.
The random 1$\sigma$-uncertainty due to errors in the measurements of 
$\beta$ is typically 0.03\,dex, and the application of Eq.~\ref{gfcal} 
yields the flux-weighted gravity
to within 0.13\,dex (systematic 
1$\sigma$-uncertainties), see Fig.~\ref{loggfloggf}.

The conversion of $\log g_\mathrm{F}^{\beta}$ to the $\beta$-based
surface gravity $\log g^{\beta}$ is facilitated by correction with the
$[c_1]$-based effective temperature, as established before.
The uncertainties of $T_\mathrm{eff}^{[c_1]}$ contribute little to the
error budget of $\log g^{\beta}$, resulting in error margins of
$\pm$0.04\,dex (1$\sigma$ random)$\pm$0.13\,dex(1$\sigma$ systematic
uncertainty).\\[3mm]
\noindent
In particular the $[c_1]$--$T_\mathrm{eff}$ (assisted by a
determination of the luminosity subclass from spectroscopy) and the
$\beta$--$\log g_\mathrm{F}$ relations derived here should facilitate to
provide useful {\em starting values of $T_\mathrm{eff}$} and $\log g$
for detailed quantitative analyses of normal BA-type supergiants.
Exceptions may be the most luminous objects (those with
pronounced H$\alpha$ P-Cygni profiles like HD\,12953), 
which as a class of their own fall out of relations given here. 

\begin{figure}
\centering
\includegraphics[width=.9\linewidth]{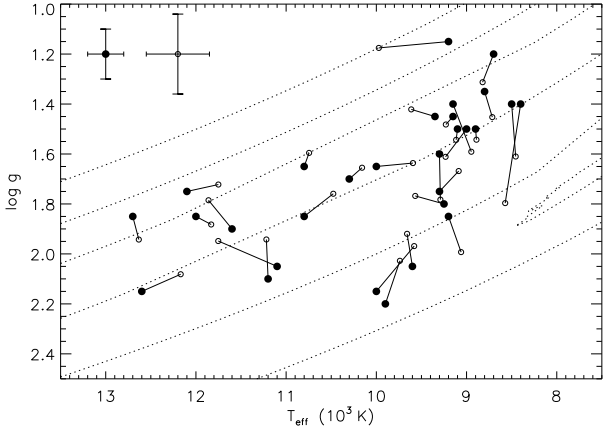}
\caption{The sample stars with Str\"omgren photometric data
in the $T_\mathrm{eff}$--$\log g$ plane. Solutions from
the spectroscopic determination are shown as dots, open circles mark
values that are obtained using photometric indicators according to our
calibrations using $[c_1]$ and $\beta$. The two corresponding solutions 
are interconnected. Typical
error bars are indicated in the upper left. Dotted lines mark
evolution tracks for rotating stars at metallicity $Z$\,=\,0.014
\citep{Ekstroem12}, from 9\,$M_{\odot}$ to 32\,$M_{\odot}$ (bottom to top).
The 9\,$M_{\odot}$ model shows a blue loop, partially displayed~here.}
\label{tefflogg}
\end{figure}

The {\em photometric starting 
values must be refined} -- preferentially through an iterative
methodology like the one discussed in Sect.~\ref{sectanalysis} -- 
in the spectroscopic analysis {\em if high
accuracy and precision is desired in the derivation of all dependent
quantities}. The reason for this is illustrated in
Fig.~\ref{tefflogg}, which shows both the results of the atmospheric
parameter determination from our detailed spectroscopic analysis and
from application of the photometric indicators discussed here in the
$T_\mathrm{eff}$--$\log g$ plane. Despite ``inconspicuous'' error bars
(see Fig.~\ref{tefflogg}), the photometric indicators without 
a subsequent refinement would yield different solutions for the
fundamental stellar parameters (mass, radius, luminosity) and 
for elemental abundances. Symptoms of the presence of such systematics would 
become evident in any more detailed analysis, as 
a failure to establish the ionization balance for multiple elements
simultaneously, and very likely a mismatch of theoretical and observed
profiles for some hydrogen lines. Results biased in such a way can easily lead
to misinterpretations and incorrect conclusions in the astrophysical
context, see Fig.~5 of \citet{SimonDiaz10} for a good example
on the effect for the elemental abundance determination, and his 
discussion of this. In terms of fundamental stellar parameters, we
see e.g.~that evolutionary masses from the comparison with stellar 
evolution tracks in Fig.~\ref{tefflogg} could be biased by up to 
$\sim$30\%.

\subsection{Tracers for studies of the ISM\label{sectISM}}
Properties of the interstellar medium (ISM) are usually investigated
by using early-type stars as background light sources, facilitating to
study absorption and scattering by the intervening interstellar material 
along the line of sight. While the topic is too broad for the scope of
the present paper, we nevertheless want to point out the usefulness of
BA-type supergiants for such studies briefly, in particular 
for the example of the ratio of total-to-selective extinction $R_V$.

\begin{figure}
\flushleft
\includegraphics[width=.95\linewidth]{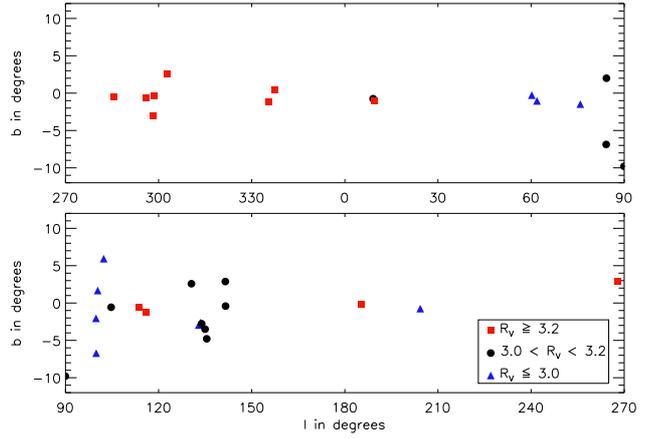}
\caption{The distribution of our sample supergiants in the Galactic plane, 
coded according to the legend for values of the 
total-to-selective extinction ratio $R_V$ along the line-of-sight.
See Sect.~\ref{sectISM} for a discussion.}
\label{rvratio}
\end{figure}

Our derived $R_V$-values for the sample stars (see Table~\ref{parameters})
cluster around the frequently quoted standard value of $R_V$\,=\,3.1 
\citep[e.g.][]{SaMa79}, with an rms scatter of 0.3. The data are displayed 
in Fig.~\ref{rvratio}, as a function of position in the Galactic plane
(Galactic coordinates $l$ and $b$ are used, note that two stars fall
outside the displayed range, HD\,34085 and HD\,87737, which are
located at higher Galactic latitude). 
We find the objects with higher values than $R_V$\,=\,3.1 concentrated in 
the southern Milky Way around 260$^{\circ}$ to 360$^{\circ}$ Galactic
longitude, and many objects with lower values in the opposite
direction, at $l$\,$\approx$\,60 to 110$^{\circ}$. This is in good agreement 
with previous findings of \cite{whittet77}.

While our sample is too small to provide any significant extension to
modern studies \citep[see e.g.][]{Winkler97,Wegner03}, which consider
hundreds of OB stars, we want to draw the attention to the potential
of BA-type supergiants for such investigations. Despite their rarity, 
they are highly useful for probing lines-of-sight towards very
distant (or highly reddened) objects in the Milky Way because of their 
much higher intrinsic visual brightness. Moreover, their usefulness for simple
investigations like presented here extends to distances even far beyond the 
Milky Way \citep[e.g.][]{Kudritzki08,Kudritzki12,U09}. It is reassuring in
this context to find the $R_V$-values for many independent Galactic 
sightlines to cluster close to the canonical value. A ratio of
total-to-selective extinction of 3.1 may therefore be a good
assumption for many extragalactic environments, where full SED
information of individual stars, required for the determination of
$R_V$, is unavailable at present. 

In addition, luminous BA-type
supergiants facilitate also more sophisticated studies of the ISM 
in other galaxies when high/intermediate-resolution spectroscopy 
becomes feasible. Examples are investigations of the neutral interstellar 
gas via the Na\,D lines (for B-type supergiants, with no stellar 
contribution to these lines) or of diffuse interstellar absorption bands 
\citep[see e.g.][]{Cordiner08a,Cordiner08b}.

\section{Summary}
A sample of 35 bright Galactic BA-type supergiants was introduced.
Quantitative non-LTE analyses of high-resolution and high-S/N spectra
were performed with the aim to provide a homogeneous set of atmospheric 
parameters at highest accuracy and precision. The study provides the
most comprehensive dataset on effective temperatures, surface
gravities, helium abundances, microturbulent, macroturbulent and
rotational velocities of Galactic BA-type supergiants so far. In
addition, the interstellar reddening and the ratio of total-to-selective 
extinction towards the sample stars were determined.

First applications of the data show that established relations from
the reference literature for BA-type supergiants are outdated. An 
improved empirical spectral-type--$T_{\rm eff}$ and $T_{\rm
eff}$--$B.C.$ scale was derived, as well as new calibrations of
intrinsic Johnson $(B-V)_0$ and Str\"omgren $(b-y)_0$ colours as a
function of effective temperature were provided. It would be desirable
to extend such studies to supergiants of earlier as well as later 
spectral types, with overall improved number statistics.

Photometric $T_{\rm eff}$-determinations 
based on the reddening-free Johnson $Q$-index were found to be of 
limited use for high-precision/accuracy studies of BA-type
supergiants because of large errors of typically
$\pm$5\%\,(1$\sigma$ statistical)$\pm$3\%\,(1$\sigma$ systematic),
compared to a spectroscopically achieved precision of 1-2\%
(combined statistical and systematic uncertainty with our
methodology). On the other hand, the reddening-free Str\"omgren 
$[c_1]$-index and Str\"omgren $\beta$ are highly promising for 
deriving good starting points for further quantitative investigations,
with uncertainties of $\pm$1\%\,$\pm$2.5\% in $T_{\rm eff}$,
and $\pm$0.04$\pm$0.13\,dex in $\log g$ \,(1$\sigma$-statistical, 
1$\sigma$-systematic,~respectively).
Finally, the potential of BA-type supergiants as tools for studying
ISM properties, also in other galaxies, was briefly addressed.

Besides the immediate objectives, this paper provides the preparatory 
work for addressing important questions of modern astrophysics.
Tight observational constraints on stellar evolution in form of
fundamental stellar parameters and light element abundances as tracers
of rotational mixing will be derived in Paper~II, and constraints 
on Galactic chemical evolution, via abundance gradients of the thin
disk, in Paper~III.

\begin{acknowledgements}
      We wish to thank U.~Heber for his interest and support of the
      project, and for useful comments on the manuscript. We thank 
      K.~Butler for providing {\sc Detail} and {\sc Surface},
      K.~Fuhrmann for observing HD\,195324, and the staff at
      Calar Alto and at ESO/La Silla for performing observations in
      the runs of 2005 and 2007, respectively.
      Some of the data presented in this paper were obtained from
      the Multimission Archive at the Space Telescope Science
      Institute (MAST). STScI is operated by the Association of
      Universities for Research in Astronomy, Inc., under NASA
      contract NAS5-26555. Support for MAST for non-HST data is
      provided by the NASA Office of Space Science via grant NAG5-7584
      and by other grants and contracts. 
      This publication makes use of data products from the Two Micron
      All Sky Survey, which is a joint project of the University of
      Massachusetts and the Infrared Processing and Analysis
      Center/California Institute of Technology, funded by the
      National Aeronautics and Space Administration and the National
      Science Foundation.
      This research has made use of
      the SIMBAD database, operated at CDS, Strasbourg, France.
      We acknowledge financial support by the 
      \emph{Deut\-sche For\-schungs\-ge\-mein\-schaft, DFG\/} project
      number PR 685/3-1. Travel to the Calar Alto Observatory/Spain in
      2001 was supported by \emph{DFG} under grant PR\,685/1-1.
\end{acknowledgements}

\bibliography{bib}

\end{document}